\documentclass[fleqn,usenatbib]{mnras}
\usepackage{newtxtext,newtxmath}
\usepackage[T1]{fontenc}
\usepackage{ae,aecompl}
\DeclareRobustCommand{\VAN}[3]{#2}
\let\VANthebibliography\thebibliography
\def\thebibliography{\DeclareRobustCommand{\VAN}[3]{##3}\VANthebibliography}
\usepackage{graphicx}
\usepackage{amsmath}
\usepackage{geometry}
\usepackage{ulem}
\usepackage{tabularx,booktabs}
\usepackage{caption}
\usepackage{dirtytalk}
\usepackage{multicol}
\usepackage{underscore}
\usepackage[english]{babel}
\usepackage{multirow}
\newcommand{\ps}{\,s$^{-1}$}	
\newcommand{\cps}{\,counts\,s$^{-1}$}	

\title[Spectro-temporal studies of GX 9$+$1] {\textit{NuSTAR} and \textit{AstroSat} observations of GX 9$+$1: Spectral and temporal studies}
\author[N. T. Thomas et al.]{Neal Titus Thomas, S. B. Gudennavar\thanks{E-mail: shivappa.b.gudennavar@christuniversity.in} and S. G. Bubbly 
\\\\
Department of Physics and Electronics, CHRIST University, Bangalore 560029, Karnataka, India\\
}
\date{Accepted 2023 August 01. Received 2023 July 26; in original form 2022 October 10}
\pubyear{2023}
\begin{document}
\label{firstpage}
\pagerange{\pageref{firstpage}--\pageref{lastpage}}
\maketitle
\begin{abstract}
We have studied the spectro-temporal properties of the neutron star low mass X-ray binary GX 9$+$1 using data from \textit{NuSTAR/FPM} and \textit{AstroSat/SXT} and \textit{LAXPC}. The hardness-intensity diagram of the source showed it to be in the soft spectral state during both observations. \textit{NuSTAR} spectral analysis yielded an inclination angle ($\theta$) $=$ 29$\substack{+3\\-4}^{\circ}$ and inner disc radius ($R_{in}$) $\leq$ 19 km. Assuming that the accretion disc was truncated at the Alfv\'en radius during the observation, the upper limit of the magnetic dipole moment ($\mu$) and the magnetic field strength ($B$) at the poles of the neutron star in GX 9$+$1 were calculated to be 1.45$\times$$10^{26}$ G cm$^3$ and 2.08$\times$$10^8$ G, respectively (for $k_A$ $=$ 1). Flux resolved spectral analysis with \textit{AstroSat} data showed the source to be in the soft spectral state ($F_{disc}$/$F_{total}$ $\sim$0.9) with a monotonic increase in mass accretion rate ($\dot{m}$) along the banana branch. The analysis also showed the presence of absorption edges at $\sim$1.9 and $\sim$2.4 keV, likely due to Si XIII and S XV, respectively. Temporal analysis with \textit{LAXPC-20} data in the 0.02 $-$ 100 Hz range revealed the presence of noise components, which could be characterized with broad Lorentzian components. 
\end{abstract}
\begin{keywords}
X-rays: binaries --- stars: neutron --- accretion, accretion discs --- X-rays: individual: GX 9$+$1
\end{keywords}
\section{Introduction}
Low mass X-ray binaries (LMXBs) are gravitationally bound binary systems consisting of a compact object (neutron star or a black hole) and a companion/donor star of mass typically lower than 1 M$_{\odot}$. In neutron star low mass X-ray binaries (NS-LMXBs), wherein the compact object is a neutron star, matter is accreted from the donor star onto the surface of the neutron star by means of Roche-lobe overflow, resulting in the production of X-rays. NS-LMXBs are categorized into persistent and transient systems, based on their long-term variabilities. While persistent NS-LMXBs are characterized by persistent luminosity in X-rays, transient NS-LMXBs linger in the quiescent state  for a significant amount of time (months to years) and occasionally undergo an outburst that lasts for weeks to months. The luminosity of transient NS-LMXBs typically increases by several orders of magnitude as it evolves from the quiescent state ($L\sim$ 10$^{32}-10^{33}$ erg s$^{-1}$) into an outburst ($L\sim$ 10$^{39}$ erg s$^{-1}$). Persistent and transient NS-LMXBs are further classified as Z and atoll sources based on their correlated spectro-temporal behaviour and the patterns traced by them in their colour-colour diagrams (CCDs) \citep{Hasinger&vanderKlis1989}. Z sources are usually very bright and sometimes radiate at Eddington limit ($L_{Edd}$). Atoll sources, on the other hand, are generally less bright ($L\lesssim$ 0.5 $L_{Edd}$). Based on their luminosity, atoll sources are further classified as bright atoll sources ($L=$ 0.3 $-$ 0.5 $L_{Edd}$) and ordinary atoll sources ($L=$ 0.01 $-$ 0.3 $L_{Edd}$) \citep{Yao2021}. Bright atoll sources are mostly located near the Galactic bulge component and exhibit a wide range of X-ray luminosities ($L\sim$ 10$^{37}$ $-$ 10$^{38}$ erg s$^{-1}$), along with irregular intensity variations and apparent lack of periodic signatures. Moreover, these sources usually trace out an elongated pattern, known as the banana branch, in their CCDs \citep{Mondal2019}. 
\\[6pt]
GX 9$+$1 (also known as 4U 1758$-$20 or X Sgr X-3) is a persistently bright atoll type NS-LMXB located in the direction of the Galactic bulge, along with GX 9$+$9, GX 3$+$1 and GX 13$+$1. It was discovered on 1965 April 25 in a survey conducted with Geiger counters on-board \textit{Aerobee-Rockets} by the Naval Research Laboratory \citep{Friedman1967}. A year later, \citet{Gursky1967} observed GX 9$+$1 using a proportional counter sensitive to X-rays in the energy range 1.5 $-$ 10.0 keV, on-board an \textit{Aerobee} rocket. Subsequently, the location of the source was confirmed to be well in the Galactic bulge ($l = 9.1^{\circ}$ and $b = 1.2^{\circ}$) by \citet{Bradt1968}. Spectro-temporal studies on GX 9$+$1 with the medium energy (ME) (1 $-$ 50 keV) instrument, on-board \textit{EXOSAT}, revealed that the source flux in the energy range 1.0 $-$ 30.0 keV varied irregularly on timescales from minutes to hours \citep{Langmeier1985}. The hardness ((3.4 $-$ 7.5 keV) \cps /(1.0 $-$ 3.2 keV) \cps) of the source showed correlated changes with its intensity (1.0 $-$ 7.5 keV \cps) and the source spectrum could be fit with a model combination consisting of a blackbody ($\mathtt{bbody}$, $kT_{bb}$ $=$ 1.16$\pm0.015$ to 1.26$\pm0.015$ keV) and a thermal bremsstrahlung ($\mathtt{bremss}$, $kT$ = 13$\pm1$ to $15\pm1$ keV) component. The \textit{EXOSAT} data of GX 9$+$1 were also analysed by \citet{White1988}, where the spectrum was fit with a blackbody ($\mathtt{bbody}$, $kT_{bb}$ $=$ 1.5 keV) and a Comptonization component ($kT_e$ $=$ 3 keV) absorbed by an equivalent hydrogen column density ($N_H$) of 2.1$\times$10$^{22}$ cm$^{-2}$. The luminosities associated with the blackbody and the Comptonization component, in the 0.1 $-$ 30 keV energy range, were estimated to be 6$\times$10$^{37}$ erg \ps and 1.7$\times$10$^{38}$ erg \ps, respectively, assuming a source distance of 10 kpc as an upper limit. Based on its correlated spectro-temporal properties from \textit{EXOSAT} data, \citet{Hasinger&vanderKlis1989} classified GX 9$+$1 to be a persistently bright atoll source. \citet{Schulz1989} showed the CCD of the source to trace out only the banana branch and that its power density spectrum (PDS) could be described by relatively strong ($\sim$3.5 per cent of root mean square (RMS)) very low frequency noise (VLFN) and weak ($\sim$2.5 per cent of RMS) high frequency noise (HFN) as compared to other atoll sources. Analyzing the \textit{RXTE} observation of the source, \citet{Wijnands1998} reported the absence of kHz quasi-periodic oscillations (QPOs) in its PDS, which are the characteristics of atoll sources tracing out the banana branch in their CCD or hardness-intensity diagram (HID). A study of the spectral evolution of GX 9$+$1, using data from \textit{INTEGRAL}, showed that two $\mathtt{compTB}$ models were required to adequately fit its broadband spectra \citep{Mainardi2010}. The first $\mathtt{compTB}$ component was used to describe the dominant part of the spectrum that was interpreted as thermal Comptonization of soft seed photons ($<$ 1 keV), likely from the accretion disc, by a 3 $-$ 5 keV corona. This component did not evolve much in terms of Comptonization efficiency. The second $\mathtt{compTB}$ component, which varied more dramatically, was used to explain Comptonization spanning from bulk plus thermal Comptonization of blackbody seed photons to the blackbody emission alone. The presence and nature of the second {$\mathtt{compTB}$} component revealed a constant blackbody contribution. The size of the blackbody-emitting region ($\sim$3.4 km) as well as the corresponding X-ray luminosity (0.2$\times$10$^{38}$ erg s$^{-1}$) were found to be nearly constant. Despite selecting spectra having different hardness, they all exhibited the high/soft state with no dramatic evolution in spectral parameters. 
On the other hand, the source has exhibited variability in its long-term X-ray lightcurves. \citet{Asai2022} investigated decades’ long-term X-ray variations in bright NS-LMXBs using high quality X-ray light curves obtained from \textit{MAXI/GSC} and \textit{RXTE/ASM}. This study revealed GX 9$+$1 to exhibit an apparent sinusoidal variation with an orbital period of $\sim$10 y in its 34 y lightcurve. It was concluded that intense irradiation by the central X-ray source may induce variation of the mass transfer rate leading to the appearance of a sinusoidal periodic variation in the X-ray intensity. 
\\[6pt]
The source distance and optical/near-infrared (NIR) companion of GX 9$+$1 were not known for a long time due to a poorly determined X-ray position, despite several attempts. \citet{Iaria2005}, using data in the 0.12 $-$ 18.0 keV energy range from the \textit{BeppoSAX} mission, estimated the source distance to be 5 kpc. This study also performed flux resolved spectroscopy and found that the source spectra could be adequately fit  with a model combination comprised of an absorbed ($N_{H}$$\sim$0.8$\times$10$^{22}$ cm$^{-2}$) blackbody ($\mathtt{bbody}$) and a Comptonization ($\mathtt{compTT}$) component, along with several absorption edges corresponding to absorption due to O VII, O VIII, Ne IX, Ar XVII, Ca XX and Fe I, and an absorption line associated with Si XIV. Recently, \citet{vandenberg2017}, based on the revised source position determined using \textit{Chandra} data, identiﬁed a 16.5$\pm$0.1 mag NIR counterpart of GX 9$+$1. This study was done using K$_s$-band images obtained with the \textit{PANIC} and \textit{FourStar} cameras on the 6.5 m \textit{Magellan Baade Telescope}. The same study used more recent maps of Galactic extinction to report a relatively small source distance of 3.7$\pm$0.3 kpc (based on the \citet{Drimmel2003} map for optical extinction $A_V$) and 2.1$\pm$0.3 kpc (based on the \citet{Schultheis2014} $E(J- K_s)$ maps), for the lower limit of $N_H$ = 8$\pm$1$\times$10$^{21}$ cm$^{-2}$. Using $N_H$ = 2.21$\pm$0.09$\times$10$^{21} A_V$ cm$^{-2}$ \citep{Guver2009}, the source distance was found to be 3.3$\pm$0.3 kpc (based on \citet{Drimmel2003} map) and 1.7$\pm$0.2 kpc (based on \citet{Schultheis2014} map). These distance values indicated that GX 9$+$1 is in front or at most on the near edge of the Galactic bulge. However, since GX 9$+$1 along with many bright X-ray binaries lie in the direction of the bulge, it is plausible that many of them are actually located in the bulge, thereby implying the source distance to have a conservative lower limit of $\sim$4 kpc. 
\\[6pt]
GX 9$+$1 was observed with the \textit{FPMA} and \textit{FPMB} detectors on-board (\textit{NuSTAR}) \citep{Harrison2013} and by the Soft X-ray Telescope (\textit{SXT}) \citep{Singh2016} and Large Area X-ray Proportional Counters (\textit{LAXPC}) \citep{Agrawal2017} instruments on-board \textit{AstroSat} on 2019 February 19 and 2019 May 2, respectively. In this study, we used these data to perform broadband spectral and temporal analysis with an objective to characterize the spectro-temporal properties of the source and constrain its physical parameters such as radius of inner accretion disc, mass accretion rate, magnetic dipole moment, magnetic field strength, thickness of the boundary layer and radius of the neutron star in the system. The procedures adopted for the reduction of data are discussed in Section \ref{sec:Observation_data_reduction}. Details of the broadband spectral and temporal analysis are presented in Section \ref{sec:spectral_temporal_analyis}. The results obtained from this study are discussed and compared with previously reported results on the source in Section \ref{sec:results_discussion}. Finally, a summary of the findings from the present study is given in Section \ref{sec:conclusions}.
\section{Observation and Data Reduction}
\label{sec:Observation_data_reduction}
This study has been carried out using data from the \textit{NuSTAR} and \textit{AstroSat} missions. GX 9$+$1 was observed by \textit{NuSTAR} for an effective exposure time of 10.54 ks on 2019 February 19 (Observation ID: 30401016002, hereafter Observation 1). \textit{NuSTAR} is an X-ray observatory equipped with two parallel focal plane modules - \textit{FPMA} and \textit{FPMB}. Each of them contains a 2$\times$2 array of CdZnTe (CZT) crystal detectors, which operate in the 3.0 $-$ 79.0 keV energy range, surrounded by CsI anti-coincidence shielding \citep{Kitaguchi2014}. The detectors have an energy resolution (full width at half maximum) of 400 eV at 10 keV and 900 eV at 79 keV and a temporal resolution of 10 $\mu$s. The source was also observed by \textit{AstroSat} with the \textit{SXT} and \textit{LAXPC} in photon counting (PC) and event analysis (EA) modes, respectively, for a total effective exposure time of 108.4 ks on 2019 May 2 (Observation ID: A05\_221T01\_9000002888, hereafter Observation 2). The details of \textit{SXT} and \textit{LAXPC} instruments are given elsewhere \citep[see for e.g.][]{Thomas2022, Thomas2023}. The details of data reduction are given in the next subsections.
\subsection{\textit{NuSTAR} data reduction}
\label{subsec:Nustar_data_reduction}
Level 1 data of Observation 1 were obtained from the HEASARC archive\footnote{\label{note2}\url{https://heasarc.gsfc.nasa.gov/docs/archive.html}} and were processed using $\mathtt{NuSTARDAS v.2.1.2}$ and $\mathtt{CALDB v.20211020}$ from $\mathtt{HEASoft v.6.31.1}$, to obtain Level 2 clean event files for both \textit{FPM} detectors, individually. Since GX 9+1 is a bright source, having a count rate $>$ 100 counts s$^{-1}$ in its 1 s binned lightcurve, the statusexpr keyword was set as "STATUS==b0000xxx00xxxxxxxx000\&\&(SHIELD==0)"\footnote{\label{note6}\url{https://heasarc.gsfc.nasa.gov/docs/nustar/nustar_faq.html}}, during the $\mathtt{nupipeline}$ run. Moreover, to filter out high background activity events from the South Atlantic Anomaly (SAA), the $\mathtt{nupipeline}$ task was run with saamode$=$optimized, saacalc$=$2 and tentacle$=$no. Source images were extracted using $\mathtt{nuproducts}$, from a circular region of radius 200$\arcsec$ centered at the focal point; whereas, the background images were extracted from a circular region of radius 150$\arcsec$ away from the source. These images were then used to generate source and background lightcurves from each detector, which were merged using the $\mathtt{lcmath}$ task. Furthermore, source and background spectra, as well as response files were generated for the purpose of spectral analysis.
\subsection{\textit{AstroSat} data reduction}
\label{subsec:Astrosat_data_reduction}
Level 1 data of Observation 2 were obtained from the \textit{AstroSat} data archives\footnote{\label{note4}\url{https://astrobrowse.issdc.gov.in/astro_archive/archive/Home.jsp}}. Level 1 data from the \textit{SXT} were processed using \textit{SXT} pipeline - AS1SXTLevel2-1.4b\footnote{\label{note3}\url{https://www.tifr.res.in/~astrosat_sxt/sxtpipeline.html}} to obtain clean event files for individual orbits of the observation. These were merged using SXT Event Merger Tool\footnote{\label{note7}\url{https://www.tifr.res.in/~astrosat_sxt/dataanalysis.html}} to generate a merged clean event file. Furthermore, the source image, lightcurves and spectra were obtained from the merged clean event file using XSELECT V2.4k. The source image was extracted from an annular region of inner and outer radii of $4^\prime$ and $12^\prime$, respectively, to reduce the effect of pile-up. The background\footnote{SkyBkg\_comb\_EL3p5\_Cl\_Rd16p0\_v01.pha} and response matrix\footnote{sxt\_pc\_mat\_g0to12.rmf} files provided by the \textit{SXT} Payload Operations Centre (POC) were used for further analysis. An off-axis Auxiliary Response File (ARF) was generated using sxt\_ARFModule\footnote{sxtARFModulev03} tool. Since the data quality below 0.8 keV suffered from instrumental noise and that above 7.0 keV was poor due to uncertainties in the response and effective area, the \textit{SXT} data were restricted to the energy range 0.8 $-$ 7.0 keV for further analysis. Level 1 data from \textit{LAXPC} were processed using LAXPCSOFT (Format A)\footnote{\url{http://astrosat-ssc.iucaa.in/laxpcData}} to obtain a Level 2 event file, Good Time Interval (GTI) files, lightcurves, source and background spectra, Response Matrix Files (RMF) and PDS. As the performance of \textit{LAXPC-10} has been poor since 2018 March 28 due to an abnormal change in its gain \citep{Antia2021} and \textit{LAXPC-30} was not operational during Observation 2, data from \textit{LAXPC-20} alone were used to carry out spectral and temporal analysis. An observation log containing the details of the \textit{NuSTAR} and \textit{AstroSat} observations used for this study is given in Table \ref{Tab:Table1}.
\begin{table*}
\centering
\caption{Observation ID, date, MJD, exposure time and average count rate, respectively of the \textit{NuSTAR} and \textit{AstroSat} observations. The average count rate is obtained from 1 s net lightcurves from \textit{LAXPC-20} and \textit{FPM} and $\sim$2.37 s net lightcurve from \textit{SXT}.}
\begin{tabular}{llll}
\hline
                                                     &                &\textit{NuSTAR}       &\textit{AstroSat}\\
\hline
Observation ID                                       &                &30401016002           &A05\_221T01\_9000002888\\ 
                                                     &                &(Observation 1)       &(Observation 2)\\
Date of observation (dd-mm-yyyy)                     &                &19-02-2019            &02-05-2019\\
MJD                                                  &                &58533                 &58605\\\\
\multirow{3}{*}{Exposure time (ks)}                  & \textit{SXT}   &                      &46.2\\
                                                     & \textit{LAXPC} &                      &62.6\\
                                                     & \textit{FPM}   &10.54                 &\\\\
\multirow{3}{*}{Average count rate (counts s$^{-1}$)} & \textit{SXT} (0.8 $-$ 7.0 keV)        &    &33.92\\
                                                     & \textit{LAXPC} (4.0 $-$ 25.0 keV)     &    &1204.20\\
                                                     & \textit{FPM}   (4.0 $-$ 25.0 keV)     &    832.22 &\\
\hline
\end{tabular}
\label{Tab:Table1}
\end{table*}
\section{Spectral and temporal studies}
\label{sec:spectral_temporal_analyis}
\subsection{Lightcurves and HID}
\label{subsec:Lightcurve_and_HID}
\subsubsection{\textit{NuSTAR} lightcurves and HID}
\label{subsubsec:Nustar_lightcurve_and_HID}
\begin{figure}
    \centering
    \includegraphics[width=\columnwidth]{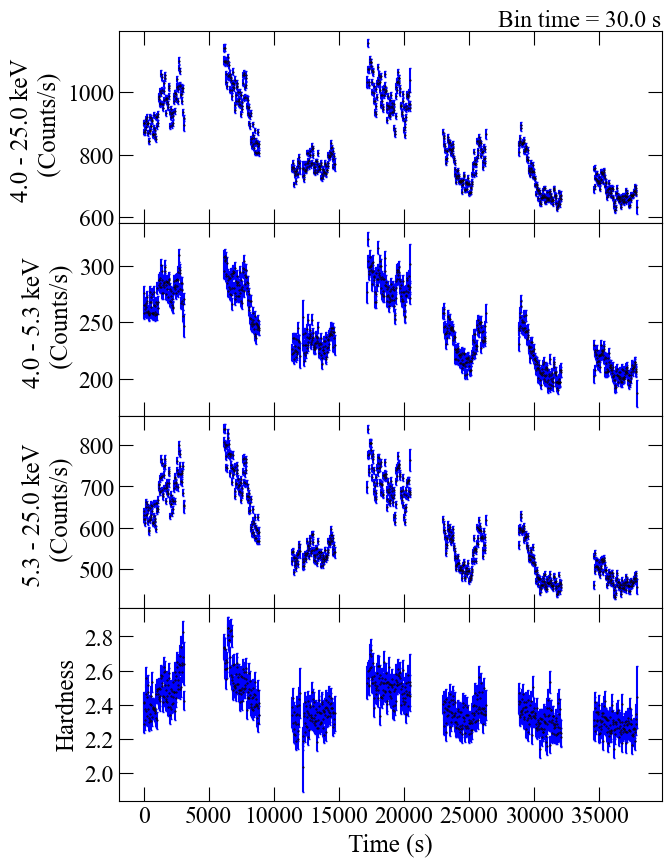}
    \caption{\textit{FPM} net lightcurves in 4.0 $-$ 25.0 (Panel 1 from the top), 4.0 $-$ 5.3 (Panel 2) and 5.3 $-$ 25.0 keV (Panel 3) energy ranges and hardness (Panel 4) as function of time.}
    \label{fig:NuSTAR_lightcurves}
\end{figure}

\begin{figure}
    \centering
    \includegraphics[width=\columnwidth]{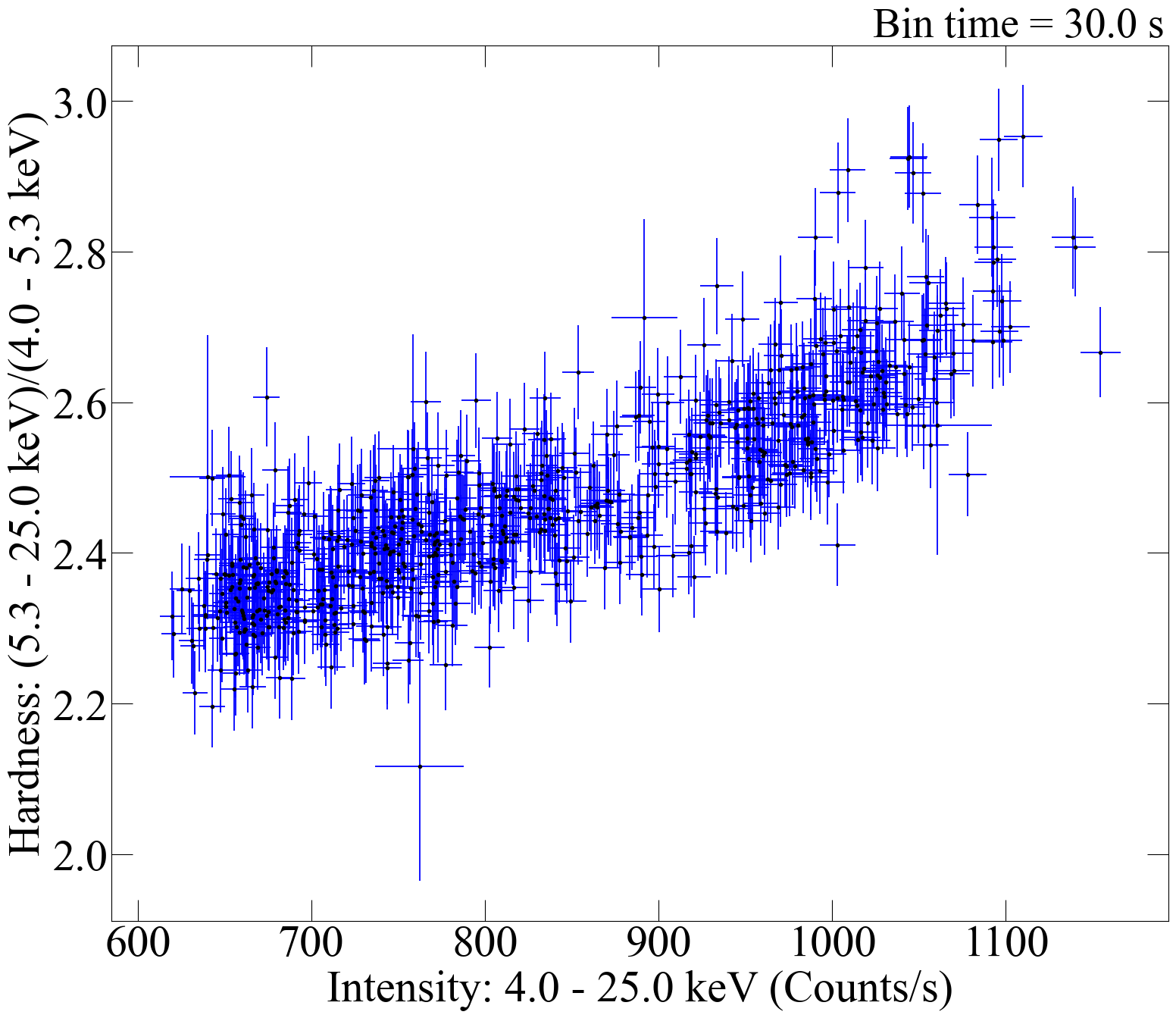}
    \caption{HID with \textit{FPM} data.}
    \label{fig:NuSTAR_HID}
\end{figure}

Net lightcurves of Observation 1 with a 30.0 s bin time were generated in the 4.0 $-$ 25.0, 4.0 $-$ 5.3 and 5.3 $-$ 25.0 keV energy ranges. These lightcurves showed the source to exhibit substantial variability (Figure \ref{fig:NuSTAR_lightcurves}). During the initial phase of the observation, the source intensity in these energy ranges showed a sharp decrease of $\sim$38, $\sim$31 and $\sim$43 per cent, respectively. This was followed by an increase in the source intensity by $\sim$63, $\sim$53 and $\sim$76 per cent and finally a gradual decrease of $\sim$45, $\sim$40 and $\sim$45 per cent towards the end of the observation. The hardness of the source, defined as the ratio of \cps in the 5.3 $-$ 25.0 keV range to that in the 4.0 $-$ 5.3 keV range, exhibited a similar trend. A 30.0 s HID was created from \textit{FPM} net lightcurves, with hardness defined as mentioned above and intensity defined as the sum of \cps in the 4.0 $-$ 25.0 keV energy range. The HID showed a positive correlation between the hardness and intensity, with hardness varying from $\sim$2.2 to $\sim$3.0 and intensity varying from $\sim$620 to $\sim$1150 \cps (Figure \ref{fig:NuSTAR_HID}). This correlation is a characteristic behaviour of atoll sources in the banana branch \citep{Hasinger&vanderKlis1989}.
\\[6pt]
\subsubsection{\textit{AstroSat} lightcurves and HIDs}
\label{subsubsec:AstroSat_lightcurves_and_HIDs}
\begin{figure}
    \centering
    \includegraphics[width=\columnwidth]{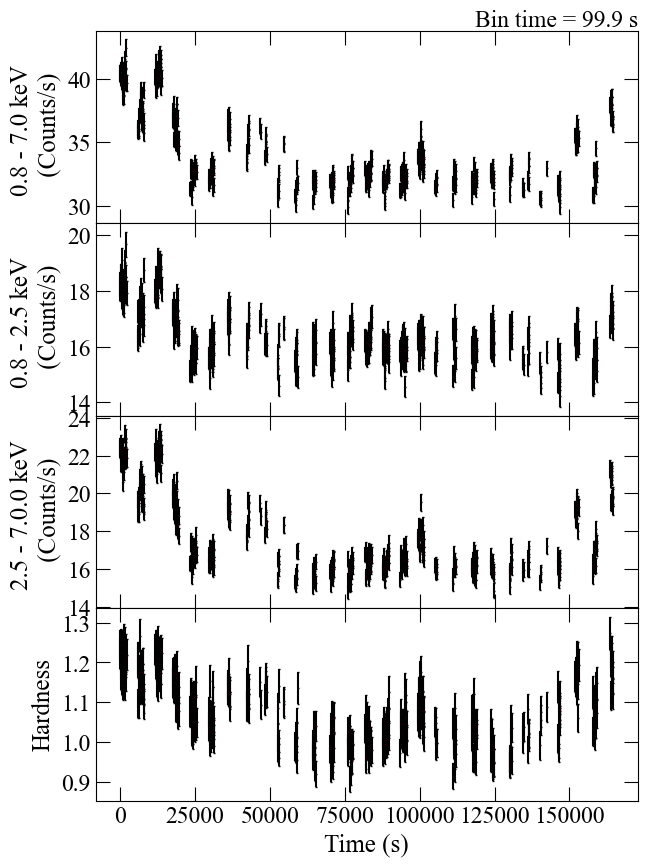}
    \caption{\textit{SXT} net lightcurve in 0.8 $-$ 7.0 (Panel 1 from the top), 0.8 $-$ 2.5 (Panel 2) and 2.5 $-$ 7.0 keV (Panel 3) energy ranges and hardness (Panel 4) as function of time.}
    \label{fig:SXT_lightcurves}
\end{figure}

\begin{figure}
    \centering
    \includegraphics[width=\columnwidth]{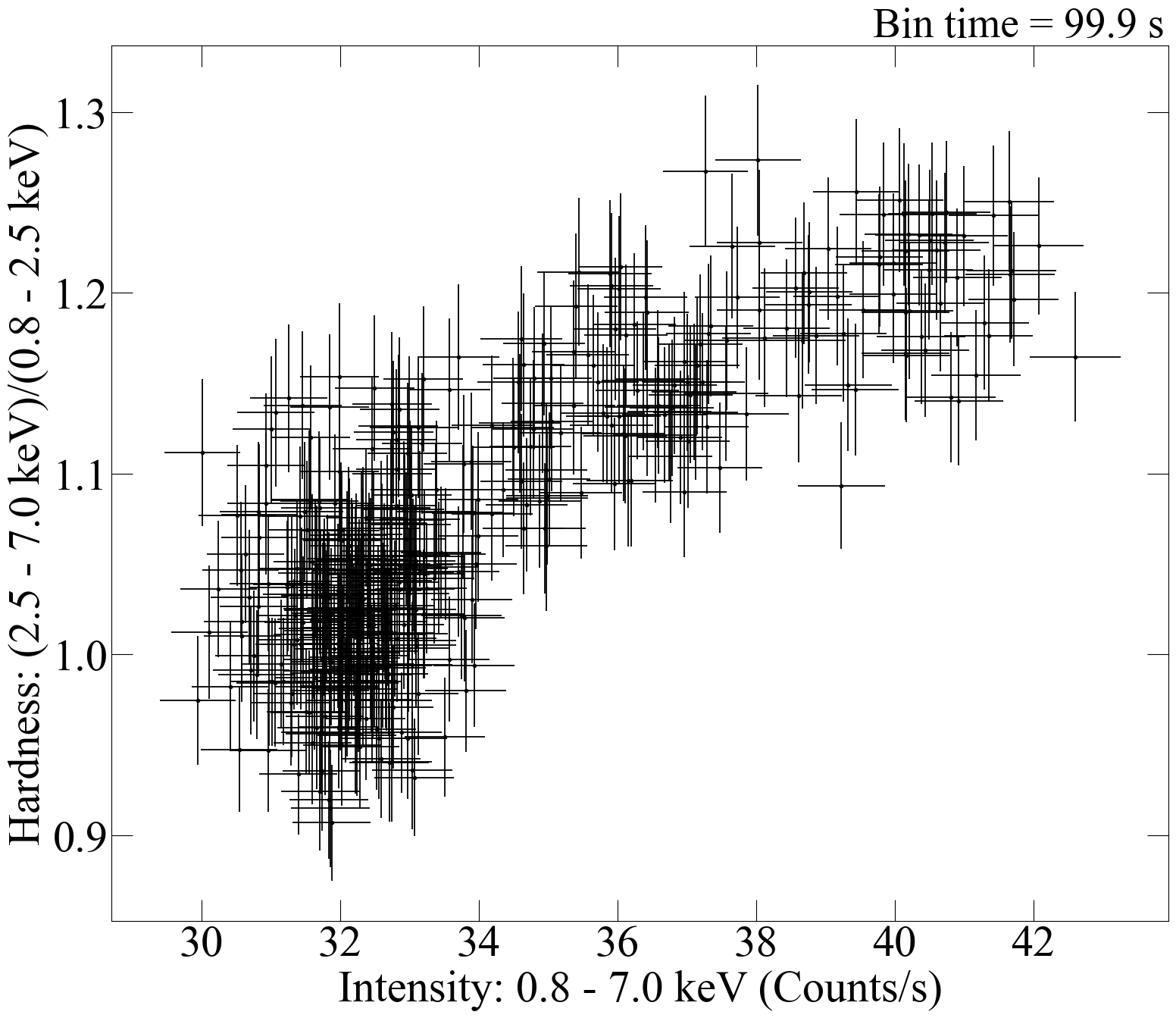}
    \caption{HID with \textit{SXT} data.}
    \label{fig:SXT_HID}
\end{figure}

\begin{figure}
    \centering
    \includegraphics[width=\columnwidth]{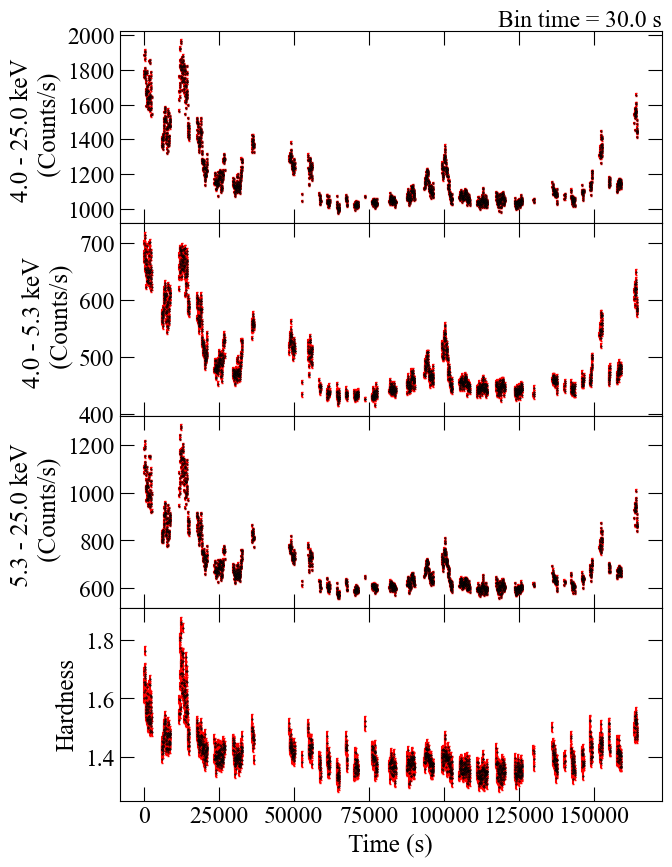}
    \caption{\textit{LAXPC-20} net lightcurves in the 4.0 $-$ 25.0 keV (Panel 1 from the top), 4.0 $-$ 5.3 keV (Panel 2) and 5.3 $-$ 25.0 keV (Panel 3) energy ranges, and hardness (Panel 4) as function of time.}
    \label{fig:LAXPC_lightcurves}
\end{figure}

\begin{figure}
    \centering
    \includegraphics[width=\columnwidth]{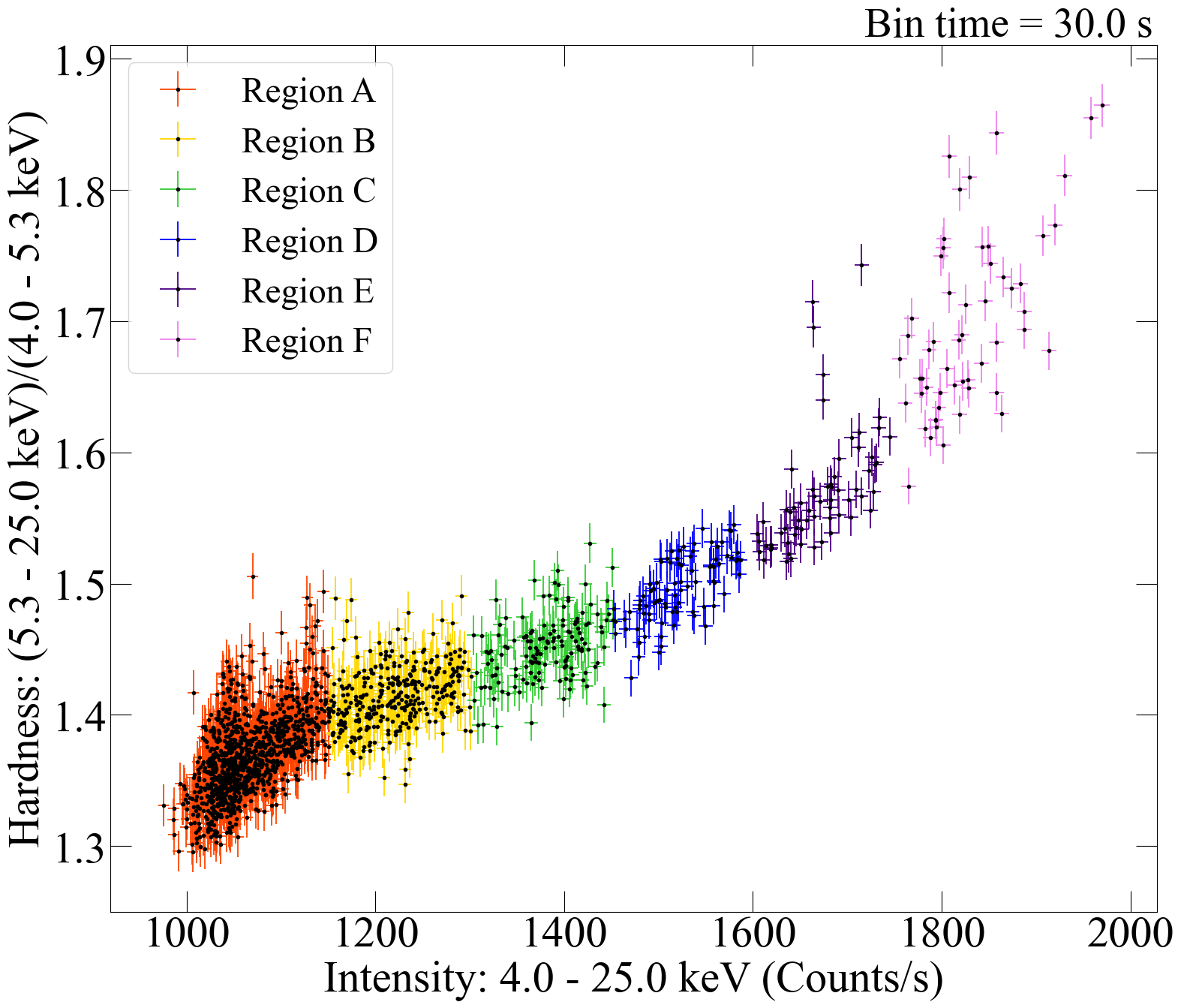}
    \caption{HID with \textit{LAXPC-20} data, resolved into different regions (A to F) based on flux to perform flux resolved spectral analysis.}
    \label{fig:LAXPC_HID}
\end{figure}

As done for Observation 1, net lightcurves of the source for Observation 2 in various energy ranges were also generated using data from both the \textit{SXT} and \textit{LAXPC-20} instruments. \textit{SXT} net lightcurves with a bin time of 99.85 s showed that the source intensity in 0.8 $-$ 7.0, 0.8 $-$ 2.5 and 2.5 $-$ 7.0 keV energy ranges decreased by $\sim$26, $\sim$21 and $\sim$32 per cent, respectively, during the initial phase of the observation; but towards the end, the intensity increased by $\sim$19, $\sim$17 and $\sim$40 per cent, respectively (Figure \ref{fig:SXT_lightcurves}). The \textit{SXT} hardness of the source, defined as the ratio of \cps in the 2.5 $-$ 7.0 keV range to the \cps in the 0.8 $-$ 2.5 keV range, followed a trend similar to that of the lightcurves i.e., an initial decrease of roughly 24 per cent followed by an increase of 27 per cent towards the end of the observation. A 99.85 s binned HID, using \textit{SXT} data, was generated by defining hardness as mentioned above  and intensity as the total \cps in the 0.8 $-$ 7.0 keV range (Figure \ref{fig:SXT_HID}). Figure \ref{fig:SXT_HID} shows a positive correlation between the hardness and intensity, with hardness varying from $\sim$0.9 to $\sim$1.3 and intensity varying from $\sim$30 to $\sim$42 \cps. Furthermore, 30 s binned net lightcurves in the same energy ranges as mentioned in Subsection \ref{subsubsec:Nustar_lightcurve_and_HID} were generated using data from \textit{LAXPC-20} (Figure \ref{fig:LAXPC_lightcurves}). These lightcurves showed that the source intensity varied substantially throughout the observation. In the beginning of the observation, an initial dip of $\sim$23 per cent in the source intensity was seen in the 4.0 $-$ 25.0, 4.0 $-$ 5.3 and 5.3 $-$ 25.0 keV ranges; followed by a gradual decrease, before finally showing an increasing trend at the end of the observation. The \textit{LAXPC-20} hardness of the source, defined as mentioned in Subsection \ref{subsubsec:Nustar_lightcurve_and_HID}, followed a trend similar to that of the \textit{LAXPC-20} lightcurves. This is also similar to the overall trend seen in the \textit{SXT} lightcurves and hardness.  However, it is to be noted that the magnitude of the increasing trend in the \textit{LAXPC-20} lightcurves and hardness is $\sim$50 per cent, which is higher than that seen in the \textit{SXT} data. As done in Subsection \ref{subsubsec:Nustar_lightcurve_and_HID}, a 30 s binned HID was created using \textit{LAXPC-20} net lightcurves (Figure \ref{fig:LAXPC_HID}). The hardness and intensity were seen to vary from roughly 1.3 to 1.8 and 1000 to 1900 \cps, respectively. Both the \textit{SXT} and \textit{LAXPC-20} HIDs (Figures \ref{fig:SXT_HID} and \ref{fig:LAXPC_HID}) reflect the positive correlation between hardness and intensity seen in the \textit{NuSTAR} HID (Figure \ref{fig:NuSTAR_HID}). This confirms that the source was in the soft spectral state during the observation and is in agreement with the previous studies on GX 9$+$1 \citep{Iaria2005, Asai2016}. 
\\[6pt]
Time-averaged and flux resolved spectral analyses were carried out to characterize the spectra of the source and to study the variation of its spectral parameters as a function of hardness and intensity. The details of these are given in the following subsections.
\subsection{Spectral analysis}
\label{subsec:Spectral_analysis}
\subsubsection{Spectral studies using \textit{NuSTAR} data}
\label{subsubsec:Spectral_studies_using_NuSTAR_data}
\begin{figure}
    \centering
    \includegraphics[width=\columnwidth]{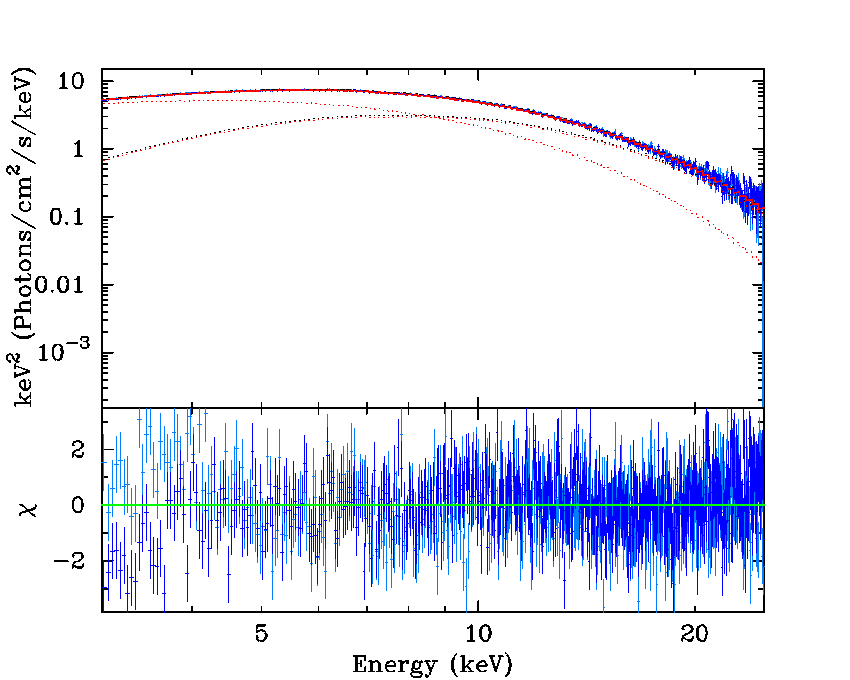}
    \caption{Time-averaged \textit{FPMA} (light blue colour) \textit{FPMB} (blue colour) unfolded spectra fit with the model combination - 
$\mathtt{constant}$ $\times$ $\mathtt{tbabs}$ $\times$ ($\mathtt{thcomp}$ $\times$ $\mathtt{discbb}$ $+$ $\mathtt{relxillNS}$)}. The residuals ($\chi$ $=$ (data $-$ model)/error) are plotted in the bottom panel.
    \label{fig:NuSTAR_spectral_fit}
\end{figure}

Broadband time-averaged spectral analysis was carried out using data from both \textit{FPM} detectors. The \textit{FPMA/B} spectra were grouped such that each spectral bin has a minimum of 25 counts and were fit simultaneously with the spectral modelling tool $\mathtt{XSPEC version: 12.13.0c}$ \citep{Arnaud1996}. The spectral fit was restricted to 3.0 $-$ 25.0 keV energy range as the source spectrum was background-dominated beyond 25.0 keV. The combined \textit{FPM} spectra were fit with the multi-colour blackbody model ($\mathtt{discbb}$) \citep{Mitsuda1984} along with the thermal Comptonization continuum model ($\mathtt{thcomp}$) \citep{Zdziarski2020}. The $\mathtt{thcomp}$ model was chosen as it agrees much better with actual Monte Carlo spectra from Comptonization than the previously prevalent $\mathtt{nthcomp}$ model \citep{Zycki1999} (see for e.g. \cite{Zdziarski2020} for details and \cite{Nied2019} for an analogous comparison of $\mathtt{thcomp}$ with $\mathtt{nthcomp}$). Absorption of source X-rays in the interstellar medium was taken into account by including the Tuebingen-Boulder Inter-Stellar Medium absorption model ($\mathtt{tbabs}$) with the solar abundance table given by \citet{Wilms2000}. The value of $N_H$ was fixed to 2.59$\times$10$^{22}$ cm$^{-2}$, obtained from the time-averaged spectral fit using \textit{AstroSat} data (Section \ref{sec:Spectral_studies_using_AstroSat_data}). This was motivated by the fact that N$_H$ peaks at lower energies and \textit{NuSTAR} spectra lacked spectral coverage $<$ 3 keV. In addition, a constant factor was multiplied to the model combination for normalization of uncertainties in the cross calibration between the two \textit{FPM} detectors. Furthermore, since $\mathtt{thcomp}$ is a convolution model, an energy binning array was supplied to the spectral fit to extend the energy range of the model combination beyond that of the \textit{FPM} detectors in the fit. The energy range 0.01 $-$ 200 keV was chosen for this purpose using the $\mathtt{Xspec}$ command $\mathtt{energies}$ $\mathtt{0.01}$ $\mathtt{200}$ $\mathtt{500}$ $\mathtt{log}$. The model combination $\mathtt{constant}$ $\times$ $\mathtt{tbabs}$ $\times$ $\mathtt{thcomp}$ $\times$ $\mathtt{discbb}$ yielded a reduced $\chi^2$/dof of 1.56/1091. There appeared to be positive residuals at $\sim$6.4 keV and above 20 keV, indicating the presence of reflected emission and a Compton hump, respectively. In order to characterise these features, the $\mathtt{relxillNS}$ model \citep{Garcia2022} was added to the model combination. This variant from the $\mathtt{relxill}$ family of models \citep{Garcia2013,Dauser2014} assumes that the accretion disc is irradiated by blackbody emission originating from the neutron star surface. The parameters of the $\mathtt{relxillNS}$ model include inner ($q_1$) and outer ($q_2$) emissivity indices, break radius ($R_{break}$) between the two emissivity indices, dimensionless spin parameter ($a$), inclination of the system ($\theta$), inner ($R_{in}$) and outer ($R_{out}$) disc radii in units of radius of Innermost Stable Circular Orbit ($R_{ISCO}$) and gravitational radius ($R_g$, where $R_g$ $=$ $GM_{NS}$/c$^2$, $G$ is the universal gravitational constant (cm$^{3}$ g$^{-1}$ s$^{-2}$) and $c$ is the speed of light (cm s$^{-1}$)), respectively, redshift ($z$), temperature of the irradiating blackbody ($kT_{bb}$), ionization parameter (log ($\xi$)), Fe abundance ($A_{Fe}$) in solar units, accretion disc density ($log (n_e)$), reflection fraction ($R_{frac}$) and normalization ($Norm$). In order to fit the spectra, a single emissivity profile was assumed ($q_1$ = $q_2$ = 3), thereby making the $R_{break}$ redundant, as done in \citet{Ludlam2022}. $R_{out}$ and $a$ were set to 990 $R_g$ and 0, respectively. As GX 9$+$1 is a galactic source, $z$ was set to 0. Furthermore, $R_{frac}$ was left to vary so that the $\mathtt{relxillNS}$ model accounts for both the irradiating blackbbody continuum and the reprocessed reflection components. Initially, Fe abundance was assumed to have solar value with $A_{Fe}$ set to 1. However, it was seen that the best fit was obtained when $A_{Fe}$ was left free and its value was constrained to be $\leq$ 0.61. The model combination $\mathtt{constant}$ $\times$ $\mathtt{tbabs}$ $\times$ ($\mathtt{thcomp}$ $\times$ $\mathtt{discbb}$ $+$ $\mathtt{relxillNS}$) resulted in an improved fit, having reduced $\chi^2$/dof of 1.20/1085. In addition, the $\mathtt{cflux}$ model was used to obtain the unabsorbed total bolometric flux ($F_{bol}$) in the 0.01 $-$ 200 keV range, the unabsorbed total flux ($F_{total}$) and unabsorbed disc flux ($F_{disc}$) in the 3.0 $-$ 25.0 keV. The unfolded fit spectra and the best-fit spectral parameters thus obtained, with their 90 per cent confidence errors are presented in Figure \ref{fig:NuSTAR_spectral_fit} and Table \ref{Tab:Table2}, respectively.
\setlength{\extrarowheight}{.35em}
\begin{table}
\centering
\caption{Best fit model spectral parameters along with their 90 per cent confidence errors obtained from fitting the time-averaged \textit{FPMA} and \textit{FPMB} spectra. In the Table, (f) - the parameter
was frozen during the fit.}
\begin{tabular}{lll}
\hline
Model & Parameter & Value \\
\hline
$\mathtt{tbabs}$                       & $N_H$ $(10^{22}$ cm$^{-2})$      & 2.59 (f)\\
$\mathtt{thcomp}$                      & $\Gamma$                         & 2.92$\substack{+0.27\\-0.30}$\\
                                       & $kT_e$ (keV)                     & 2.20$\substack{+0.09\\-0.16}$\\
                                       & $Cov_{frac}$                     & $\geq$0.87\\
$\mathtt{discbb}$                      & $kT_{in}$ (keV)                  & 1.49$\substack{+0.09\\-0.11}$\\
                                       & $Norm$                           & 125$\substack{+45\\-17}$\\
$\mathtt{relxillNS}$                   & $\theta$ ($^{\circ}$)            & 29$\substack{+3\\-4}$\\
                                       & $R_{in}$ ($R_{ISCO}$)            & $\leq$1.54\\
                                       & $kT_{bb}$ (keV)                  & 2.42$\substack{+0.01\\-0.02}$\\
                                       & $log (\xi)$ (erg cm s$^{-1}$)    & $\leq$3.30\\
                                       & $A_{Fe}$                         & $\leq$0.61\\
                                       & $log (n_e)$ (cm$^{-3}$)          & $\geq$18.47\\
                                       & $R_{frac}$                       & $\geq$3.17\\
                                       & $Norm$ ($10^{-3}$)               & 1.57$\substack{+0.20\\-0.12}$\\
$\mathtt{cflux}$                  & $F_{bol}$ ($10^{-8}$ erg cm$^{-2}$ s$^{-1}$)    & 2.31$\substack{+0.01\\-0.18}$\\
                                  & $F_{total}$ ($10^{-8}$ erg cm$^{-2}$ s$^{-1}$)  & 1.55$\substack{+0.01\\-0.20}$\\
                                  & $F_{disc}$ ($10^{-8}$ erg cm$^{-2}$ s$^{-1}$)   & 1.14$\substack{+0.03\\-0.04}$\\
                                  & $F_{disc}$/$F_{total}$                          & 0.73\\
\hline
                                  &Reduced $\chi^2/$dof                             & 1.20/1091\\
\hline
\end{tabular}
\label{Tab:Table2}
\end{table}
The best fit spectral parameters were used to compute physical properties like inner disc radius ($R_{in}$), radius of the boundary layer ($R_{max} - R_*$) and mass accretion rate ($\dot{m}$) of the source; as well as the magnetic dipole moment ($\mu$) and magnetic field strength ($B$) of the neutron star in the system. The apparent inner disc radius, $r_{in}$ (km) was computed from the normalization ($Norm$) of the $\mathtt{discbb}$ model using equation (\ref{discbb_norm}),
\begin{equation}
\begin{aligned}
Norm = \biggl(\frac{r_{in}}{D_{10}}\biggl)^2 cos \theta
\end{aligned}
\label{discbb_norm}
\end{equation}
where, $D_{10}$ is the distance to the source (in units of 10 kpc) and $\theta$ is the disc inclination angle ($^{\circ}$) of the source. The absolute or true inner disc radius $R_{in}$ (km) is related to the apparent inner disc radius $r_{in}$ as $R_{in}$ $\simeq$ $\kappa^2$ $\times$ $\xi$ $\times$ $r_{in}$, where $\kappa$ is the ratio of color temperature to effective temperature or the spectral hardening factor \citep{Shimura1995} and $\xi$ is the inner boundary correction factor \citep{Kubota1998}. $R_{in}$ was computed assuming $D_{10}$ $=$ 0.5 \citep{Iaria2005}, $\theta$ $=$ 29$^{\circ}$ (Table \ref{Tab:Table2}), $\kappa$ $=$ 2.0 \citep{Shimura1995} and $\xi$ $=$ 0.412 \citep{Kubota1998}. The upper limit of $R_{in}$ was found to be 12 km. On the other hand, the $\mathtt{relxillNS}$ model yielded an inner disc radius $R_{in}$ $\leq$ 1.54 $R_{ISCO}$ (where, $R_{ISCO}$ $=$ 6 $R_g$), which corresponds to $\leq$ 9.24 $R_g$ $=$ 19 km. This implies that the accretion disc is truncated very close to the neutron star in the system. As done by \citet{Saavedra2023}, assuming, $R_{in}$ obtained from the $\mathtt{relxillNS}$ model to be the Alfv\'en radius, where magnetic field pressure is balanced by the ram pressure from infalling material, thereby causing truncation of the accretion disc; the upper limit of the magnetic dipole moment, $\mu$ (G cm$^3$) was computed with equation (\ref{equation2}) \citep{Cackett2009},
\begin{equation}
\begin{aligned}
\mu = 3.5\times 10^{23} D_{3.5} (M_{NS}^{\dagger})^2 \left(\frac{x}{k_{A}}\right)^{7/4} \left(\frac{f_{ang}\times F_{bol}^{\dagger}}{\eta}\right)^{1/2} 
\end{aligned}
\label{equation2}
\end{equation}
where,  $D_{3.5}$ is the distance to the source in units of 3.5 kpc, $M_{NS}^{\dagger}$ is the mass of the neutron star in units of 1.4 M$_{\odot}$, $x$ is given by $R_{in}$ $=$ $xGM/c^2$ 
\citep{Cackett2009}, $k_A$ is a coefficient that depends on the conversion from spherical to disc accretion, $f_{ang}$ is the
anisotropy correction factor, $F_{bol}^{\dagger}$ is the total bolometric flux in units of $10^{-9}$ erg cm$^{-2}$ s$^{-1}$ and $\eta$ is the accretion efficiency in the Schwarzschild metric. Assuming $M_{NS}^{\dagger}$ $=$ 1, $D_{3.5}$ $=$ 1.43, $k_A$ $=$ 1, $f_{ang}$ $=$ 1and $\eta$ $=$ 0.1 for a neutron star of radius $R_{NS}$ $=$ 11.2 km \citep{Steiner2018}, the upper limit of $\mu$ and the magnetic field strength ($B$) at the poles of the neutron star in GX 9$+$1 were calculated to be 1.45$\times$$10^{26}$ G cm$^3$ and 2.08$\times$$10^8$ G, respectively. Furthermore, assuming $k_A$ $=$ 0.5, the upper limit of $\mu$ and $B$ at the poles were calculated to be 4.9$\times 10^{26}$ G cm$^3$ and 7.0$\times$$10^8$ G, respectively. Using the $F_{bol}$ in the 0.01 $-$ 200 keV, the mass accretion rate $\dot{m}$ (g cm$^{-2}$ s$^{-1}$ and $M_{\odot}$ y$^{-1}$) at the surface of neutron star in GX 9$+$1 was estimated using equation  (\ref{equation1}) \citep{Galloway2008},
\begin{equation}
\begin{aligned}
\dot{m} = \frac{6.7\times 10^3\times F_{bol}^{\dagger}\left(1+z\right) D_{10}^2}{M_{NS}^{\dagger}\times R_{NS}^{\dagger}} 
\end{aligned}
\label{equation1}
\end{equation}

where, $F_{bol}^{\dagger}$ is the total bolometric flux in units of $10^{-9}$ erg cm$^{-2}$ s$^{-1}$, $M_{NS}^{\dagger}$ and $R_{NS}^{\dagger}$ are the mass and radius of the neutron star in units of 1.4 M$_{\odot}$ and 10 km, respectively. The surface redshift $z$ is given by, $1+z =\left(1-\frac{2GM_{NS}}{R_{NS}c^2}\right)^{-1/2}$, where $G$ is the universal gravitational constant (cm$^{3}$ g$^{-1}$ s$^{-2}$) and $c$ is the speed of light (cm s$^{-1}$), $M_{NS}$ and $R_{NS}$ are the mass and radius of the neutron star in the system, respectively. This estimation of mass accretion rate was made with the assumptions, $D_{10}$ $=$ 0.5 \citep{Iaria2005}, $M_{NS}$ = 1.4 M$_{\odot}$ and $R_{NS}$ $=$ 11.2 km \citep{Steiner2018} (hence, $M_{NS}^{\dagger}$ $=$ 1, $R_{NS}^{\dagger}$ $=$ 1.12). Its value was found to be 3.32$\substack{+0.01\\-0.26}$$\times$$10^4$ g cm$^{-2}$ s$^{-1}$ or 8.30$\substack{+0.04\\-0.65}$$\times$$10^{-9}$ $M_{\odot}$ y$^{-1}$.
\\[6pt]
In a thin accretion disc, matter rotates in its innermost region at approximately the Keplerian velocity. If this velocity is more than the rotational velocity of the accreting neutron star, a \say{boundary layer} is formed, where most of the gravitational potential energy released in the accretion process is converted into the rotational energy of the accreting matter. Furthermore, since this energy originates from a small region close to the neutron star, the boundary layer is thought to be hotter than the accretion disc and produce higher energy radiation. The Comptonization of seed photons may be caused due to such a boundary layer \citep{Revnivtsev2006}. The thickness of the boundary layer ($R_{max} - R_{*}$) was calculated using equation (\ref{Equation4}) \citep{Popham2001},
\begin{equation}
\begin{aligned}
log (R_{max} - R_{*}) \simeq 5.02 + 0.245[log(\dot{m}_{-9.85})]^{2.19}
\end{aligned}
\label{Equation4}
\end{equation}
where, $R_{max}$ is the maximal radial extent of the boundary layer, $R_{*}$ is the radius of neutron star and $\dot{m}_{-9.85}$ is the mass accretion rate in units of $10^{-9.85}$ M$_{\odot}$ y$^{-1}$.
\subsubsection{Spectral studies using \textit{AstroSat} data}
\label{sec:Spectral_studies_using_AstroSat_data}
\begin{figure}
\centering
\includegraphics[width=\columnwidth]{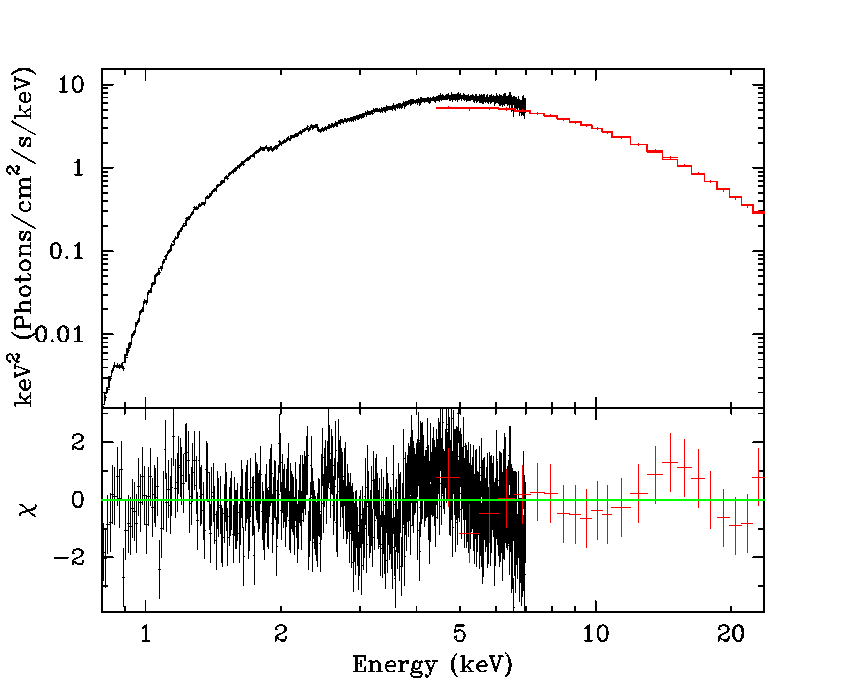}
\caption{Time-averaged \textit{SXT} and \textit{LAXP-20} unfolded spectra fit with the model combination - 
$\mathtt{constant}$ $\times$ $\mathtt{tbabs}$ $\times$ $\mathtt{edge}$ $\times$ $\mathtt{edge}$ $\times$ $\mathtt{thcomp}$ $\times$ $\mathtt{discbb}$. The residuals ($\chi$ $=$ (data $-$ model)/error) are plotted in the bottom panel.}
\label{fig:SXT_LAXPC_spectral_fit}
\end{figure}

\begin{figure}
\centering
\includegraphics[width=\columnwidth]{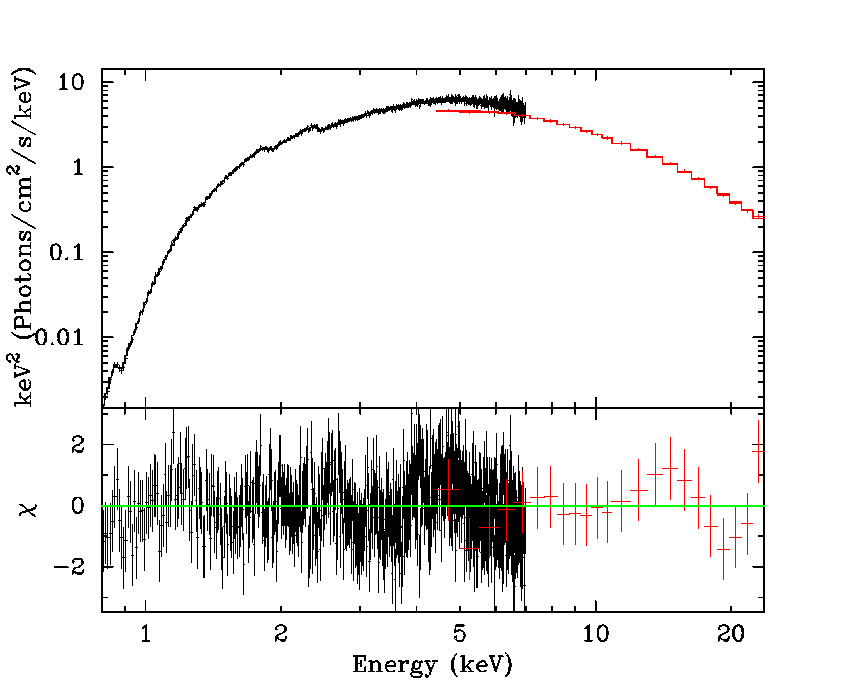}
\caption{\textit{SXT} and \textit{LAXPC-20} unfolded spectra of Region A, fit with the model combination - 
$\mathtt{constant}$ $\times$ $\mathtt{tbabs}$ $\times$ $\mathtt{edge}$ $\times$ $\mathtt{edge}$ $\times$ $\mathtt{thcomp}$ $\times$ $\mathtt{discbb}$. The residuals ($\chi$ $=$ (data $-$ model)/error) are plotted in the bottom panel.}
\label{fig:RegionA_SXT_LAXPC_spectral_fit}
\end{figure}

The simultaneous broadband spectral capabilities of \textit{AstroSat} with the \textit{SXT} and \textit{LAXPC} instruments were used to perform spectral studies in the 0.8 $-$ 25.0 keV energy range. Time-averaged broadband spectral studies were conducted by first generating simultaneous GTIs from both the \textit{SXT} and \textit{LAXPC-20} instruments for the entire duration of the observation. These GTIs were used to generate \textit{SXT} and \textit{LAXPC-20} spectra. The \textit{SXT} spectrum was grouped such that each bin contained a minimum of 30 counts and, following \citet{Misra2017}, the \textit{LAXPC-20} spectrum was grouped such that the energy bins were $\sim$6 per cent of the mean energy. The \textit{SXT} and \textit{LAXPC-20} simultaneous grouped spectra were fit jointly in the 0.8 $-$ 25.0 keV energy range with the spectral modelling tool $\mathtt{XSPEC version: 12.13.0c}$ \citep{Arnaud1996}. The spectral fit was restricted to 0.8 $-$ 7.0 keV for the \textit{SXT} instrument, due to uncertainties in its response and effective area; and to 4.0 $-$ 25.0 keV for the \textit{LAXPC-20} instrument, due to presence of instrumental noise below 4.0 keV and large uncertainties in the background around 30 keV owing to K X-ray fluorescence of Xe \citep{Antia2017}. The \textit{SXT} and \textit{LAXPC-20} spectra were jointly fit with the model combination - $\mathtt{constant}$ $\times$ $\mathtt{tbabs}$ $\times$ $\mathtt{thcomp}$ $\times$ $\mathtt{discbb}$, as done for Observation 1 in Section \ref{subsubsec:Spectral_studies_using_NuSTAR_data}. The constant factor here accounts for the uncertainties in the cross calibration between \textit{SXT} and \textit{LAXPC-20} instruments. A systematic error of 3 per cent was added as prescribed by \citet{Bhattacharya2017}. Finally, a gain fit was performed with the slope of the gain fixed to unity and the offset left to vary to account for the non-linear change in the detector gain of the \textit{SXT} instrument. Furthermore, as done in Section \ref{subsubsec:Spectral_studies_using_NuSTAR_data}, an energy binning array was supplied to the spectral fit with the $\mathtt{Xspec}$ command $\mathtt{energies}~\mathtt{0.01}~\mathtt{200}~\mathtt{500}~\mathtt{log}$, to extend the energy range of the model combination beyond that of \textit{SXT} and \textit{LAXPC-20} in the fit. This spectral fit yielded reduced $\chi^2$/dof of 2.05/634. The $\mathtt{discbb}$ model was replaced with the blackbody model - $\mathtt{bbodyrad}$, making the model combination - $\mathtt{constant}$ $\times$ $\mathtt{tbabs}$ $\times$ $\mathtt{thcomp}$ $\times$ $\mathtt{bbodyrad}$, which yielded reduced $\chi^2$/dof of 2.8/634. In both these model combinations, positive residuals were observed around 2 keV. In order to address this, the above model combination was modified to include both $\mathtt{discbb}$ and $\mathtt{bbodyrad}$ models. However, this did not improve the spectral fit and resulted in a very low inner disc temperature (7.8$\times$$10^{-4}$ keV) and a very high value of $\mathtt{discbb}$ $Norm$ (8.6$\times$$10^{14}$). Upon close inspection, we deemed the positive residuals around 2 keV to be an indication of the presence of absorption lines in the \textit{SXT} spectrum. Several NS-LMXBs have been reported to exhibit absorption edges in their spectra \citep[see for e.g.][]{Schulz2016}. These features could be associated with the presence of ionized matter around the system. In order to address the observed absorption features, the standard $\mathtt{edge}$ function was used, making the final model combination - $\mathtt{constant}$ $\times$ $\mathtt{tbabs}$ $\times$ $\mathtt{edge}$ $\times$ $\mathtt{edge}$ $\times$ $\mathtt{thcomp}$ $\times$ $\mathtt{discbb}$, which resulted in a good fit having $\chi^2$/dof of 1.07/630. In addition, the $\mathtt{cflux}$ model was used to obtain the unabsorbed total bolometric flux ($F_{bol}$) in the 0.01 $-$ 200 keV range, the unabsorbed total flux ($F_{total}$) and unabsorbed disc flux ($F_{disc}$) in the 0.8 $-$ 25.0 keV. The unfolded fit spectra and the best-fit spectral parameters thus obtained with their 90 per cent confidence errors are presented in Figure \ref{fig:SXT_LAXPC_spectral_fit} and Table \ref{Tab:Table3}, respectively.
\\[6pt]
In order to study spectral parameters of the source in different flux states, flux-resolved spectral analysis was carried out by resolving the \textit{LAXPC-20} HID into six regions (A, B, C, D, E and F) based on the net flux in the 4.0 $-$ 25.0 keV energy range (Figure \ref{fig:LAXPC_HID}). The details of flux ranges used to flux-resolve the \textit{LAXPC-20} HID are given in Table \ref{Tab:Table4}. Following the procedure adopted to fit the time-averaged spectra, simultaneous, grouped \textit{SXT} and \textit{LAXPC-20} spectra of the source from all the six regions were fit jointly with the model combination - $\mathtt{constant}$ $\times$ $\mathtt{tbabs}$ $\times$ $\mathtt{edge}$ $\times$ $\mathtt{edge}$ $\times$ $\mathtt{thcomp}$ $\times$ $\mathtt{discbb}$. This yielded a good fit with an average reduced $\chi^2$ of 0.99. The best-fit spectral parameters thus obtained with their 90 per cent confidence errors are presented in Table \ref{Tab:Table5}. A representative fit spectra of Region A is given in Figure \ref{fig:RegionA_SXT_LAXPC_spectral_fit}. 
\begin{table}
\centering
\caption{Best fit model spectral parameters along with their 90 per cent confidence errors obtained from fitting the time-averaged \textit{SXT} and \textit{LAXPC-20} spectra.}
\begin{tabular}{lll}
\hline
Model & Parameter & Value \\ 
\hline
$\mathtt{tbabs}$                       & $N_H$ $(10^{22}$ cm$^{-2})$      & 2.59$\pm$0.04\\
$\mathtt{edge}$                        & $Line$ (keV)                     & 1.88$\pm$0.02\\
                                       & $\tau$                           & 0.16$\substack{+0.04\\-0.02}$\\
$\mathtt{edge}$                        & $Line$ (keV)                     & 2.42$\substack{+0.02\\-0.01}$\\
                                       & $\tau$                           & 0.24$\substack{+0.04\\-0.03}$\\
$\mathtt{thcomp}$                      & $\Gamma$                         & $\leq$2.12\\
                                       & $kT_e$ (keV)                     & 3.13$\substack{+0.38\\-0.22}$\\
                                       & $Cov_{frac}$                     & 0.06$\substack{+0.12\\-0.04}$\\
$\mathtt{discbb}$                      & $kT_{in}$ (keV)                  & 1.88$\substack{+0.06\\-0.05}$\\
                                       & $Norm$                           & 66$\pm$7\\
$\mathtt{cflux}$                       & $F_{bol}$ ($10^{-8}$ erg cm$^{-2}$ s$^{-1}$)   & 1.87$\pm$0.03\\
                                       & $F_{total}$ ($10^{-8}$ erg cm$^{-2}$ s$^{-1}$) & 1.70$\pm$0.03\\
                                       & $F_{disc}$ ($10^{-8}$ erg cm$^{-2}$ s$^{-1}$)  & 1.54$\substack{+0.07\\-0.06}$\\
                                       & $F_{disc}$/$F_{total}$                         & 0.91\\
\hline
                                       &Reduced $\chi^2/$dof                            & 1.07/630\\
\hline
\end{tabular}
\label{Tab:Table3}
\end{table}
\begin{table}
\caption{Details of flux ranges in the 4.0 $-$ 25.0 keV energy range used for flux-resolved spectral analysis.}
\begin{tabular}{lll}
\hline
Region & Flux range (\cps) & Average hardness \\
\hline
Region A & 900 $-$ 1150   & 1.37 \\
Region B & 1150 $-$ 1300  & 1.41 \\ 
Region C & 1300 $-$ 1450  & 1.45 \\
Region D & 1450 $-$ 1600  & 1.50 \\   
Region E & 1600 $-$ 1750  & 1.57 \\   
Region F & 1750 $-$ 2000  & 1.70 \\   
\hline
\end{tabular}
\label{Tab:Table4}
\end{table}
\begin{table*}
\caption{Best fit model spectral parameters along with their 90 per cent confidence errors obtained from fitting the flux resolved \textit{SXT} and \textit{LAXPC-20} spectra.}
\begin{tabular}{llllllll}
\hline
Model & Parameters & Region A & Region B & Region C & Region D & Region E & Region F\\
\hline
$\mathtt{tbabs}$ & $N_H$ $(10^{22}$ cm$^{-2})$ & 2.53$\pm$0.05 & 2.56$\substack{+0.08\\-0.07}$ & 2.69$\pm$0.04 & 2.75$\pm$0.07 & 2.74$\pm$0.07 & 2.60$\substack{+0.13\\-0.07}$\\
$\mathtt{edge}$ & $Line$ (keV) & 1.88$\pm$0.02 & 1.89$\pm$0.03 & 1.89$\pm$0.03 & 1.9$\pm$0.03 & 1.88$\pm$0.04 & 1.83$\pm$0.03\\
& $\tau$ & 0.15$\substack{+0.05\\-0.03}$ & 0.19$\substack{+0.07\\-0.05}$ & 0.16$\substack{+0.04\\-0.03}$ & 0.19$\substack{+0.05\\-0.03}$ & 0.20$\pm$0.04 & 0.22$\substack{+0.17\\-0.05}$\\
$\mathtt{edge}$ & $Line$ (keV) & 2.42$\pm$0.02 & 2.42$\substack{+0.03\\-0.04}$ & 2.42$\pm$0.02 & 2.44$\pm$0.02 & 2.39$\substack{+0.05\\-0.04}$ & 2.45$\substack{+0.04\\-0.05}$\\
& $\tau$ & 0.24$\substack{+0.06\\-0.04}$ & 0.26$\substack{+0.06\\-0.05}$ & 0.26$\substack{+0.04\\-0.03}$ & 0.28$\substack{+0.06\\-0.04}$ & 0.27$\pm$0.05 & 0.25$\substack{+0.14\\-0.05}$\\
$\mathtt{thcomp}$ & $\Gamma$ & 1.63$\substack{+0.69\\-1.63}$ & 2.98$\substack{+0.35\\-0.49}$ & 1.86$\substack{+0.23\\-0.17}$ & 1.87$\substack{+0.64\\-0.69}$ & 2.08$\substack{+0.46\\-0.28}$ & 1.69$\substack{+0.44\\-0.26}$\\
& $kT_e$ (keV) & 3.02$\substack{+0.45\\-0.25}$ & 4.61$\substack{+1.02\\-0.52}$ & 3.70$\substack{+0.55\\-0.35}$ & 3.62$\substack{+0.65\\-0.40}$ & 3.68$\substack{+1.50\\-0.53}$ & 3.44$\substack{+1.36\\-0.50}$\\
& $Cov_{frac}$ & 0.10$\substack{+0.08\\-0.06}$ & 0.28$\substack{+0.17\\-0.14}$ & 0.07$\pm$0.02 & 0.08$\substack{+0.03\\-0.02}$ & 0.12$\substack{+0.06\\-0.05}$ & 0.07$\substack{+0.36\\-0.06}$\\
$\mathtt{discbb}$ & $kT_{in}$ (keV) & 1.73$\substack{+0.08\\-0.06}$ & 1.9$\pm$0.07 & 2.02$\substack{+0.06\\-0.07}$ & 2.06$\substack{+0.08\\-0.09}$ & 2.06$\substack{+0.08\\-0.10}$ & 2.19$\substack{+0.10\\-0.19}$\\
& $Norm$ & 80$\pm$11 & 61$\substack{+11\\-9}$ & 55$\substack{+8\\-7}$ & 57$\substack{+11\\-8}$ & 60$\substack{+12\\-9}$ & 52$\substack{+22\\-8}$\\
\hline
&Reduced $\chi^2/$dof & 0.93/627 & 1.02/598 & 0.99/586 & 0.96/559 & 1.04/552 & 1.0/544\\
\hline
\end{tabular}
\label{Tab:Table5}
\end{table*}
As done with the time-averaged spectral fit, the $\mathtt{cflux}$ model was used to obtain $F_{bol}$, $F_{total}$ and $F_{disc}$. The mass accretion rate ($\dot{m}$) for each region was computed using equation (\ref{equation1}), as done for Observation 1. The values of $F_{bol}$, $F_{total}$, $F_{disc}$, $F_{disc}/F_{total}$ and $\dot {m}$ for all flux resolved regions are presented in Table \ref{Tab:Table6}. 
\begin{table*}
\centering
\caption{{$F_{bol}$, $F_{total}$, $F_{disc}$, $F_{disc}$/$F_{total}$ and $\dot{m}$ values derived from the best fit parameters obtained from flux-resolved spectral analysis.}}
\begin{tabular}{lllllll}
\hline
Region & $F_{bol}$ & $F_{total}$ & $F_{disc}$ & $F_{disc}$/$F_{total}$ & $\dot{m}$ & $\dot{m}$ \\ 
& ($10^{-8}$ erg cm$^{-2}$ s$^{-1}$) & ($10^{-8}$ erg cm$^{-2}$ s$^{-1}$) & ($10^{-8}$ erg cm$^{-2}$ s$^{-1}$) & &($10^4$ g cm$^{-2}$ s$^{-1}$) & ($10^{-9}$ $M_{\odot}$ y$^{-1}$)\\
\hline
Region A & 1.67$\pm$0.03 & 1.51$\pm$0.03 & 1.32$\substack{+0.08\\-0.05}$ & 0.87 & 2.40$\pm$0.04 & 6.00$\pm$0.11\\
Region B & 1.78$\pm$0.04 & 1.62$\pm$0.03 & 1.46$\pm$0.04 & 0.90 & 2.56$\pm$0.06 &  6.40$\pm$0.14\\
Region C & 2.00$\pm$0.04 & 1.83$\pm$0.04 & 1.73$\substack{+0.05\\-0.06}$ & 0.95 & 2.87$\pm$0.06 & 7.19$\pm$0.14\\
Region D & 2.15$\pm$0.06 & 2.00$\pm$0.05 & 1.90$\substack{+0.03\\-0.06}$ & 0.95 & 3.09$\pm$0.09 & 7.73$\pm$0.22\\
Region E & 2.42$\substack{+0.06\\-0.05}$ & 2.22$\pm$0.05 & 2.04$\substack{+0.05\\-0.06}$ & 0.92 & 3.48$\substack{+0.09\\-0.07}$ &  8.70$\substack{+0.22\\-0.18}$\\
Region F & 2.58$\substack{+0.07\\-0.06}$ & 2.38$\substack{+0.06\\-0.05}$ & 2.22$\substack{+0.1\\-0.15}$ & 0.93 & 3.71$\substack{+0.1\\-0.09}$ &  9.27$\substack{+0.25\\-0.22}$\\
\hline
\label{Tab:Table6}
\end{tabular}
\end{table*}
\\[6pt] 
In the subsection below, details of the temporal analysis carried out are presented.
\subsection{Temporal analysis}
\label{subsec:Temporal_analysis}
\begin{figure}
    \centering
    \includegraphics[width=\columnwidth]{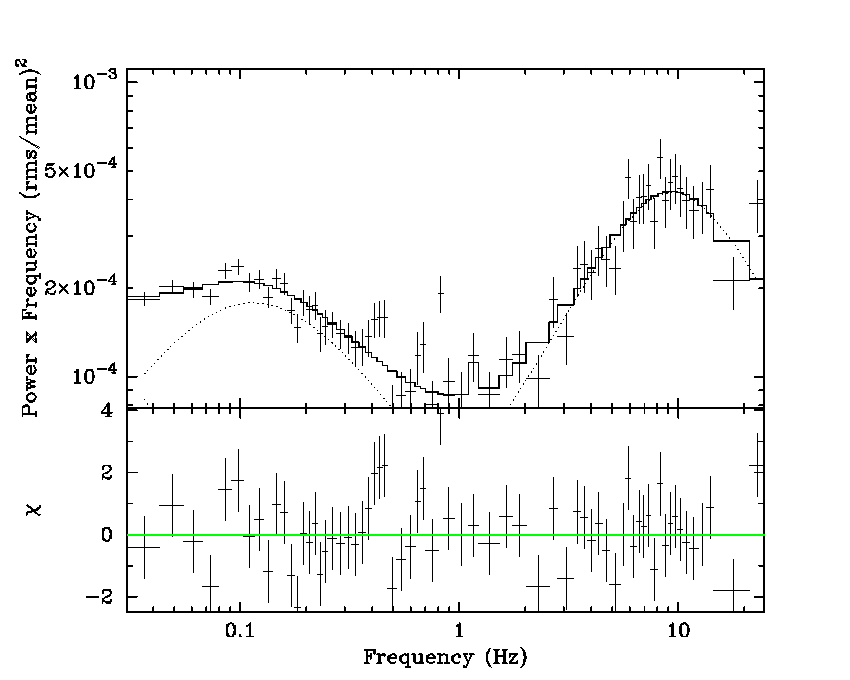}
    \caption{0.02 $-$ 40 Hz PDS, using \textit{LAXPC-20} data, fit with Lorentzians. The residuals ($\chi = (data-model)/error$) are plotted in the bottom panel.}
    \label{fig:PDS}
\end{figure}

Temporal analysis was carried out, using only \textit{LAXPC-20} data (as the \textit{NuSTAR} data suffered from deadtime), in order to investigate the temporal properties of the source such as broadband noise and QPOs. A time-averaged, RMS normalized PDS in the frequency range 0.02 $-$ 100 Hz was generated in the 4.0 $-$ 25.0 keV energy range, using data from \textit{LAXPC-20}. The PDS was modelled with Lorentzians to characterize the temporal features exhibited by the source. The temporal fit was restricted to 0.02 $-$ 40 Hz due to the absence of significant power above 40 Hz. The PDS could be adequately fit with two Lorentzian components, with the line energy of one of the Lorentzian components fixed at zero and all other parameters left free (Figure \ref{fig:PDS}). The PDS did not show the presence of narrow QPO features. The best-ﬁt parameters along with their 90 per cent confidence errors of time-averaged temporal analysis are presented in Table \ref{Tab:Table7}. In addition, energy dependent time-lag of the $\sim$4.65 Hz Lorentzian component was investigated using the methods discussed in \citet{Nowak1999}. The time-averaged cross-spectrum was used to compute time-lag at 0.01, 0.1 and 1 Hz, in several energy ranges between 3.0 $-$ 80.0 keV, using the 3.0 $-$ 3.9 keV range as reference. However, no trend could be discerned from the plot of time-lag as function of energy due to the large errors on time-lag. 
\\[6pt]
\begin{table}
\centering
\caption{Best fit model parameters of PDS. In the Table, (f) - the parameter was frozen during the fit and * - the error of that parameter could not be constrained.}
\begin{tabular}{lll}
\hline
Model & Parameters & Value \\ 
\hline
$\mathtt{Lorentzian}$ $\mathtt{1}$ & $Line$ (Hz) & 0 (f) \\
& $Width$ (Hz) & 0.17$\pm$0.01 \\
& $Norm$ ($10^{-4}$)  & 7.02$\substack{+0.23\\-0.22}$ \\
$\mathtt{Lorentzian}$ $\mathtt{2}$ & $Line$ (Hz) & 4.65$\substack{+1.07\\-1.62}$ \\
& $Width$ (Hz) & 16.80$\substack{+*\\-2.84}$ \\
& $Norm$ ($10^{-3}$)& 1.01$\substack{+0.10\\-0.01}$ \\
\hline
&Reduced $\chi^2$ /dof &1.59/62\\
\hline
\end{tabular}
\label{Tab:Table7}
\end{table}
\section{Results and Discussion}
\label{sec:results_discussion}
In this work, broadband spectral and temporal analysis of the atoll NS-LMXB GX 9$+$1 has been carried out using data from \textit{FPM} detectors on-board \textit{NuSTAR} and \textit{SXT} and \textit{LAXPC-20} on-board \textit{AstroSat}. To the best of our knowledge, systematic studies on these datasets have not been reported before. The \textit{FPM} (Figure \ref{fig:NuSTAR_lightcurves}), \textit{SXT} (Figure \ref{fig:SXT_lightcurves}) and \textit{LAXPC-20} lightcurves (Figure \ref{fig:LAXPC_lightcurves}) show that during both Observations 1 and 2, the source exhibited substantial variations in intensity and hardness. The \textit{FPM} (Figure \ref{fig:NuSTAR_HID}), \textit{SXT} (Figure \ref{fig:SXT_HID}) and \textit{LAXPC-20} (Figure \ref{fig:LAXPC_HID}) HIDs show a positive correlation between the hardness and intensity indicating that the source was in the banana branch of atoll track during both the observations. This behaviour was previously seen in a \textit{BeppoSAX} campaign of the source, where the source exhibited similar behaviour in its CCD \citep{Iaria2005}.
\\[6pt]
The broadband spectra of GX 9$+$1, in the 3.0 $-$ 25.0 keV range for Observation 1 and 0.8 $-$ 25.0 keV range for Observation 2, were used to investigate its spectral properties and calculate the physical parameters of the system. Time-averaged \textit{FPM} spectra could be adequately modelled using model combination consisting of an absorbed ($\mathtt{tbabs}$) multitemperature blackbody model ($\mathtt{discbb}$) and a Comptonization model ($\mathtt{thcomp}$) along with the addition of a reflection model ($\mathtt{relxillNS}$). The photon index ($\Gamma$ $=$ 2.92$\substack{+0.27\\-0.30}$) obtained from the time-averaged spectral fit shows the source to be in the soft spectral state during Observation 1. This is supported by the value of inner disc radius $R_{in}$ $\leq$ 19 km and $R_{in}$ $\leq$ 12 km, obtained from the $\mathtt{relxillNS}$ and $\mathtt{discbb}$ models, respectively, indicating that the accretion disc is truncated very close to the neutron star in the system; and also by the value of $F_{disc}/F_{total} = 0.73$, indicating that the source spectrum was disc dominated. It is to be noted that there is some discrepancy between the values of $R_{in}$ obtained from the $\mathtt{relxillNS}$ and $\mathtt{discbb}$ models. This could possibly be due to the choice of correction factors $\kappa$ and $\xi$ used to calculate $R_{in}$ from the $\mathtt{discbb}$ model. The value of $A_{Fe}$ parameter was found to be $\leq$0.61, indicating the source to have a subsolar Fe abundance. A similar behaviour ($A_{Fe}$ $<$ 0.7) has been previously reported in the NS-LMXB, Cygnus X-2 \citep{Mondal2018}. The reflection component in Observation 1, could be due to irradiation of the accretion disc by a boundary layer, giving rise to an Fe emission line profile around 6.4 keV and a Compton hump above 20 keV. The thickness of the boundary layer ($R_{max} - R_*$) was calculated to be 7.5 km. Using this with the $R_{in}$ value (19 km) obtained from the $\mathtt{relxillNS}$ model, the radius of the neutron star in GX 9$+$1 was constrained to $\leq$ 11.5 km. Assuming that the inner accretion disc is truncated at the Alfv\'en radius, the upper limit of the magnetic dipole moment and the magnetic field strength at the poles of the neutron star in GX 9$+$1 were calculated to be 1.45$\times$$10^{26}$ G cm$^3$ and 2.08$\times$$10^8$ G, respectively, for $k_A$ $=$ 1; and for $k_A$ $=$ 0.5, their values were found to be 4.9$\times$$10^{26}$ G cm$^3$ and 7.0$\times$$10^8$ G, respectively. It is important to note that these calculations are based on the strong assumption of magnetic truncation of the accretion disc, wherein the accretion disc is truncated due to the balance of magnetic field pressure of the neutron star in the system and the ram pressure of infalling material. Furthermore, the inclination of the source was found to be 29$\substack{+3\\-4}^{\circ}$.To the best of our knowledge, there has not been any previous estimation of the magnetic dipole moment, magnetic field strength of the neutron star in GX 9$+$1 and the inclination angle of the system. Hence the results obtained in this work could not be compared or confirmed with literature. Time-averaged spectral fit of Observation 2, with \textit{SXT} and \textit{LAXPC-20} data, showed that the 0.8 $-$ 25.0 keV spectra could be adequately fit with the model combination - $\mathtt{constant}$ $\times$ $\mathtt{tbabs}$ $\times$ $\mathtt{edge}$ $\times$ $\mathtt{edge}$ $\times$ $\mathtt{thcomp}$ $\times$ $\mathtt{discbb}$. The value of $F_{disc}$/$F_{total}$ $=$ 0.91, obtained from the spectral fit, confirms the source to be in the soft spectral state during Observation 2, as well. The values of spectral parameters such as $kT_e$ from the $\mathtt{thcomp}$ model, $kT_{in}$ and $Norm$ from the $\mathtt{discbb}$ model are fairly consistent in both Observations 1 and 2 (Tables \ref{Tab:Table2}, \ref{Tab:Table3}). This substantiates that the source was in the soft spectral state during both observations. However, it is to be noted that the value of covering fraction ($Cov_{frac}$) shows large variations between the two observations indicating that the source exhibits variability within the same spectral state. Flux-resolved spectroscopy of GX 9$+$1 shows that the $\mathtt{discbb}$ $Norm$ exhibits a gradual decrease from $\sim$80 in the lower banana branch to $\sim$52 in the upper banana branch. In addition, the bolometric flux and mass accretion rate show a monotonic increasing trend along the banana branch (Table \ref{Tab:Table6}). Two absorption edges, due to the presence of ionized material around the system, are found in both the time-averaged and flux-resolved spectra. Their line values of $\sim$1.9 and $\sim$2.4 keV likely correspond to absorption edges of Si XIII and S XV, respectively \citep{Foster2012}. Although several absorption features have been previously reported from GX 9$+$1 \citep{Iaria2005}, none of them are compatible with the ones reported in this study.
\\[6pt]
Temporal analysis was carried out on the 0.02 $-$ 100 Hz RMS normalized PDS in the 4.0 $-$ 25.0 keV energy range, using \textit{LAXPC-20} data to probe the temporal properties of the source. The RMS normalized PDS was dominated by Poisson noise and showed the presence of broadband noise components, which were fit with Lorentzians. A zero centered and $\sim$4.65 Hz Lorentzian component could fit the PDS adequately. No narrow QPO features were detected. Investigation of the energy dependent time-lag of the $\sim$4.65 Hz broadband noise component, at 0.01, 0.1 and 1 Hz, did not yield any substantial results due to large error estimates on the time-lags. 
\section{Conclusions}
\label{sec:conclusions}
The NS-LMXB source GX 9$+$1 was studied using data from \textit{FPM} detectors on-board \textit{NuSTAR} and \textit{SXT} and \textit{LAXPC-20} instruments on-board \textit{AstroSat}. The source exhibited substantial variations in its various lightcurves in the 0.8 $-$ 25 keV energy range (Figures \ref{fig:NuSTAR_lightcurves}, \ref{fig:SXT_lightcurves} and \ref{fig:LAXPC_lightcurves}). The HIDs of the source (Figure \ref{fig:NuSTAR_HID}, \ref{fig:SXT_HID} and \ref{fig:LAXPC_HID}) clearly indicated that it was in the banana branch during both the observations. Broadband spectral analysis of Observation 1 showed that the time-averaged spectrum of the source could be adequately modelled with an absorbed ($\mathtt{tbabs}$) multitemperature blackbody model ($\mathtt{discbb}$) and a Comptonization model ($\mathtt{thcomp}$) along with the addition of a reflection model ($\mathtt{relxillNS}$) (Figure \ref{fig:NuSTAR_spectral_fit}, Table \ref{Tab:Table2}). This spectral fit yielded an inner disc radius, $R_{in}$ $\leq$ 19 km and disc inclination angle, $\theta$ $=$ 29$\substack{+3\\-4}^{\circ}$. Assuming that the accretion disc was truncated at the Alfv\'en radius during the observation, the upper limit of the magnetic dipole moment and the magnetic field strength at the poles of the neutron star in GX 9$+$1 were calculated to be 1.45$\times$$10^{26}$ G cm$^3$ and 2.08$\times$$10^8$ G, respectively, for $k_A$ $=$ 1; and for $k_A$ $=$ 0.5, their values were found to be 4.9$\times$$10^{26}$ G cm$^3$ and 7.0$\times$$10^8$ G, respectively. Using the calculated boundary layer thickness ($\simeq$ 7.5 km) and $R_{in}$ (19 km) obtained from the $\mathtt{relxillNS}$ model, the radius of the neutron star in GX 9$+$1 was constrained to $\leq$ 11.5 km. Spectral analysis of Observation 2 showed that the 0.8 $-$ 25.0 keV spectra of the source could be adequately fit with a model combination consisting of absorbed ($\mathtt{tbabs}$) multitemperature blackbody model ($\mathtt{discbb}$), a Comptonization model ($\mathtt{thcomp}$) along with two edge models ($\mathtt{edge}$) (Figure \ref{fig:SXT_LAXPC_spectral_fit}, Table \ref{Tab:Table3}). The time-averaged spectral fits of both Observations 1 and 2, showed that the values of spectral parameters such as $kT_e$, $kT_{in}$, $\mathtt{discbb}$ $Norm$ and $F_{disc}/F_{total}$ are fairly consistent; and that the source was in the soft spectral state state during both observations. Two absorption edges were found to be present in both the time-averaged and flux-resolved spectra of Observation 2. Their line values of $\sim$1.9 and $\sim$2.4 keV likely correspond to the absorption edges of Si XIII and S XV, respectively.
\\[6pt]
Temporal analysis carried out using \textit{LAXPC-20} data, in the frequency and energy ranges of 0.02 $-$ 100 Hz and 4.0 $-$ 25.0 keV, respectively, showed the presence of only broadband noise components, with no narrow QPO features. Investigation of the energy dependent time-lag of the $\sim$4.65 Hz broadband noise component, at 0.01, 0.1 and 1 Hz, yielded negative results. 

\section*{Acknowledgements}
We thank immensely the anonymous reviewer for his/her insightful comments/suggestions which significantly enriched the content of the manuscript. The authors thank the \textit{SXT} and \textit{LAXPC} POC teams at Tata Institute of Fundamental Research (TIFR), Mumbai, India for the timely release of data and providing the necessary software tools. This publication uses data from the \textit{AstroSat} mission of the Indian Space Research Organisation (ISRO), archived at the Indian Space Science Data Centre (ISSDC) and \textit{NuSTAR} mission of the National Aeronautics and Space Administration. This work has made use of software provided by HEASARC. The authors acknowledge the financial support (No. DS-2B-13013(2)/9/2019-Sec.2 dated 2019 April 29) of ISRO under \textit{AstroSat} Archival Data Utilization Program. One of the authors (SBG) thanks the Inter-University Centre for Astronomy and Astrophysics (IUCAA), Pune, India for the Visiting Associateship. 
\section*{Data availability}
The data utilised in this article are available at \textit{AstroSat}$-$ISSDC website (\url{http://astrobrowse.issdc.gov.in/astro_archive/archive/Home.jsp)}. The software used for data analysis is available at HEASARC website (\url{https://heasarc.gsfc.nasa.gov/lheasoft/download.html)}.
\section*{ORCID IDs}
Neal Titus Thomas: https://orcid.org/0000-0001-9460-3264
\newline Shivappa B. Gudennavar: https://orcid.org/0000-0002-9019-9441
\newline S. G. Bubbly: https://orcid.org/0000-0003-1234-0662

\bibliography{bibliography}{}

\begin{thebibliography}{}
\makeatletter
\relax
\def\mn@urlcharsother{\let\do\@makeother \do\$\do\&\do\#\do\^\do\_\do\%\do\~}
\def\mn@doi{\begingroup\mn@urlcharsother \@ifnextchar [ {\mn@doi@}
  {\mn@doi@[]}}
\def\mn@doi@[#1]#2{\def\@tempa{#1}\ifx\@tempa\@empty \href
  {http://dx.doi.org/#2} {doi:#2}\else \href {http://dx.doi.org/#2} {#1}\fi
  \endgroup}
\def\mn@eprint#1#2{\mn@eprint@#1:#2::\@nil}
\def\mn@eprint@arXiv#1{\href {http://arxiv.org/abs/#1} {{\tt arXiv:#1}}}
\def\mn@eprint@dblp#1{\href {http://dblp.uni-trier.de/rec/bibtex/#1.xml}
  {dblp:#1}}
\def\mn@eprint@#1:#2:#3:#4\@nil{\def\@tempa {#1}\def\@tempb {#2}\def\@tempc
  {#3}\ifx \@tempc \@empty \let \@tempc \@tempb \let \@tempb \@tempa \fi \ifx
  \@tempb \@empty \def\@tempb {arXiv}\fi \@ifundefined
  {mn@eprint@\@tempb}{\@tempb:\@tempc}{\expandafter \expandafter \csname
  mn@eprint@\@tempb\endcsname \expandafter{\@tempc}}}

\bibitem[\protect\citeauthoryear{{Agrawal}}{{Agrawal}}{2017}]{Agrawal2017}
{Agrawal} P.~C.,  2017, \mn@doi [JA\&A] {10.1007/s12036-017-9449-6}, \href
  {https://ui.adsabs.harvard.edu/abs/2017JApA...38...27A} {38, 27}

\bibitem[\protect\citeauthoryear{Antia et~al.,}{Antia et~al.}{2017}]{Antia2017}
Antia H.~M.,  et~al., 2017, \mn@doi [\apjs] {10.3847/1538-4365/aa7a0e}, \href
  {https://ui.adsabs.harvard.edu/abs/2017ApJS..231...10A} {231, 10}

\bibitem[\protect\citeauthoryear{{Antia} et~al.,}{{Antia}
  et~al.}{2021}]{Antia2021}
{Antia} H.~M.,  et~al., 2021, \mn@doi [JA\&A] {10.1007/s12036-021-09712-8},
  \href {https://ui.adsabs.harvard.edu/abs/2021JApA...42...32A} {42, 32}

\bibitem[\protect\citeauthoryear{{Arnaud}}{{Arnaud}}{1996}]{Arnaud1996}
{Arnaud} K.~A.,  1996, in {Jacoby} G.~H.,  {Barnes} J.,  eds,  Astronomical
  Society of the Pacific Conference Series Vol. 101, Astronomical Data Analysis
  Software and Systems V. p.~17

\bibitem[\protect\citeauthoryear{{Asai}, {Mihara}, {Mastuoka}  \&
  {Sugizaki}}{{Asai} et~al.}{2016}]{Asai2016}
{Asai} K.,  {Mihara} T.,  {Mastuoka} M.,   {Sugizaki} M.,  2016, \mn@doi
  [\pasj] {10.1093/pasj/psw048}, \href
  {https://ui.adsabs.harvard.edu/abs/2016PASJ...68...50A} {68, 50}

\bibitem[\protect\citeauthoryear{{Asai}, {Mihara}  \& {Matsuoka}}{{Asai}
  et~al.}{2022}]{Asai2022}
{Asai} K.,  {Mihara} T.,   {Matsuoka} M.,  2022, \mn@doi [\pasj]
  {10.1093/pasj/psac049}, \href
  {https://ui.adsabs.harvard.edu/abs/2022PASJ...74..974A} {74, 974}

\bibitem[\protect\citeauthoryear{{Bhattacharya}}{{Bhattacharya}}{2017}]{Bhattacharya2017}
{Bhattacharya} D.,  2017, \mn@doi [JA\&A] {10.1007/s12036-017-9461-x}, \href
  {https://ui.adsabs.harvard.edu/abs/2017JApA...38...51B} {38, 51}

\bibitem[\protect\citeauthoryear{{Bradt}, {Naranan}, {Rappaport}  \&
  {Spada}}{{Bradt} et~al.}{1968}]{Bradt1968}
{Bradt} H.,  {Naranan} S.,  {Rappaport} S.,   {Spada} G.,  1968, \mn@doi [\apj]
  {10.1086/149613}, \href
  {https://ui.adsabs.harvard.edu/abs/1968ApJ...152.1005B} {152, 1005}

\bibitem[\protect\citeauthoryear{{Cackett}, {Altamirano}, {Patruno}, {Miller},
  {Reynolds}, {Linares}  \& {Wijnands}}{{Cackett} et~al.}{2009}]{Cackett2009}
{Cackett} E.~M.,  {Altamirano} D.,  {Patruno} A.,  {Miller} J.~M.,  {Reynolds}
  M.,  {Linares} M.,   {Wijnands} R.,  2009, \mn@doi [\apjl]
  {10.1088/0004-637X/694/1/L21}, \href
  {https://ui.adsabs.harvard.edu/abs/2009ApJ...694L..21C} {694, L21}

\bibitem[\protect\citeauthoryear{{Dauser}, {Garcia}, {Parker}, {Fabian}  \&
  {Wilms}}{{Dauser} et~al.}{2014}]{Dauser2014}
{Dauser} T.,  {Garcia} J.,  {Parker} M.~L.,  {Fabian} A.~C.,   {Wilms} J.,
  2014, \mn@doi [\mnras] {10.1093/mnrasl/slu125}, \href
  {https://ui.adsabs.harvard.edu/abs/2014MNRAS.444L.100D} {444, L100}

\bibitem[\protect\citeauthoryear{{Drimmel}, {Cabrera-Lavers}  \&
  {L\'opez-Corredoira}}{{Drimmel} et~al.}{2003}]{Drimmel2003}
{Drimmel} R.,  {Cabrera-Lavers} A.,   {L\'opez-Corredoira} M.,  2003, \mn@doi
  [A\&A] {10.1051/0004-6361:20031070}, \href
  {https://doi.org/10.1051/0004-6361:20031070} {409, 205}

\bibitem[\protect\citeauthoryear{{Foster}, {Ji}, {Smith}  \&
  {Brickhouse}}{{Foster} et~al.}{2012}]{Foster2012}
{Foster} A.~R.,  {Ji} L.,  {Smith} R.~K.,   {Brickhouse} N.~S.,  2012, \mn@doi
  [\apj] {10.1088/0004-637X/756/2/128}, \href
  {https://ui.adsabs.harvard.edu/abs/2012ApJ...756..128F} {756, 128}

\bibitem[\protect\citeauthoryear{{Friedman}, {Byram}  \& {Chubb}}{{Friedman}
  et~al.}{1967}]{Friedman1967}
{Friedman} H.,  {Byram} E.~T.,   {Chubb} T.~A.,  1967, \mn@doi [Science]
  {10.1126/science.156.3773.374}, \href
  {https://ui.adsabs.harvard.edu/abs/1967Sci...156..374F} {156, 374}

\bibitem[\protect\citeauthoryear{{Galloway}, {Muno}, {Hartman}, {Psaltis}  \&
  {Chakrabarty}}{{Galloway} et~al.}{2008}]{Galloway2008}
{Galloway} D.~K.,  {Muno} M.~P.,  {Hartman} J.~M.,  {Psaltis} D.,
  {Chakrabarty} D.,  2008, \mn@doi [A\&ASS] {10.1086/592044}, \href
  {https://ui.adsabs.harvard.edu/abs/2008ApJS..179..360G} {179, 360}

\bibitem[\protect\citeauthoryear{{Garc{\'\i}a}, {Dauser}, {Reynolds},
  {Kallman}, {McClintock}, {Wilms}  \& {Eikmann}}{{Garc{\'\i}a}
  et~al.}{2013}]{Garcia2013}
{Garc{\'\i}a} J.,  {Dauser} T.,  {Reynolds} C.~S.,  {Kallman} T.~R.,
  {McClintock} J.~E.,  {Wilms} J.,   {Eikmann} W.,  2013, \mn@doi [\apj]
  {10.1088/0004-637X/768/2/146}, \href
  {https://ui.adsabs.harvard.edu/abs/2013ApJ...768..146G} {768, 146}

\bibitem[\protect\citeauthoryear{{Garc{\'\i}a}, {Dauser}, {Ludlam}, {Parker},
  {Fabian}, {Harrison}  \& {Wilms}}{{Garc{\'\i}a} et~al.}{2022}]{Garcia2022}
{Garc{\'\i}a} J.~A.,  {Dauser} T.,  {Ludlam} R.,  {Parker} M.,  {Fabian} A.,
  {Harrison} F.~A.,   {Wilms} J.,  2022, \mn@doi [\apj]
  {10.3847/1538-4357/ac3cb7}, \href
  {https://ui.adsabs.harvard.edu/abs/2022ApJ...926...13G} {926, 13}

\bibitem[\protect\citeauthoryear{{Gursky}, {Gorenstein}  \&
  {Giacconi}}{{Gursky} et~al.}{1967}]{Gursky1967}
{Gursky} H.,  {Gorenstein} P.,   {Giacconi} R.,  1967, \mn@doi [\apj]
  {10.1086/180097}, \href
  {https://ui.adsabs.harvard.edu/abs/1967ApJ...150L..75G} {150, L75}

\bibitem[\protect\citeauthoryear{{G{\"u}ver} \& {{\"O}zel}}{{G{\"u}ver} \&
  {{\"O}zel}}{2009}]{Guver2009}
{G{\"u}ver} T.,  {{\"O}zel} F.,  2009, \mn@doi [\mnras]
  {10.1111/j.1365-2966.2009.15598.x}, \href
  {https://ui.adsabs.harvard.edu/abs/2009MNRAS.400.2050G} {400, 2050}

\bibitem[\protect\citeauthoryear{{Harrison} et~al.,}{{Harrison}
  et~al.}{2013}]{Harrison2013}
{Harrison} F.~A.,  et~al., 2013, \mn@doi [\apj] {10.1088/0004-637X/770/2/103},
  \href {https://ui.adsabs.harvard.edu/abs/2013ApJ...770..103H} {770, 103}

\bibitem[\protect\citeauthoryear{{Hasinger} \& {van der Klis}}{{Hasinger} \&
  {van der Klis}}{1989}]{Hasinger&vanderKlis1989}
{Hasinger} G.,  {van der Klis} M.,  1989, A\&A, \href
  {https://ui.adsabs.harvard.edu/abs/1989A&A...225...79H} {225, 79}

\bibitem[\protect\citeauthoryear{{Iaria}, {di Salvo}, {Robba}, {Lavagetto},
  {Burderi}, {Stella}  \& {van der Klis}}{{Iaria} et~al.}{2005}]{Iaria2005}
{Iaria} R.,  {di Salvo} T.,  {Robba} N.~R.,  {Lavagetto} G.,  {Burderi} L.,
  {Stella} L.,   {van der Klis} M.,  2005, \mn@doi [A\&A]
  {10.1051/0004-6361:20042231}, \href
  {https://ui.adsabs.harvard.edu/abs/2005A&A...439..575I} {439, 575}

\bibitem[\protect\citeauthoryear{{Kitaguchi} et~al.,}{{Kitaguchi}
  et~al.}{2014}]{Kitaguchi2014}
{Kitaguchi} T.,  et~al., 2014, Proceedings of SPIE, \href
  {https://ui.adsabs.harvard.edu/abs/2014SPIE.9144E..1RK} {9144, 91441-R}

\bibitem[\protect\citeauthoryear{{Kubota}, {Tanaka}, {Makishima}, {Ueda},
  {Dotani}, {Inoue}  \& {Yamaoka}}{{Kubota} et~al.}{1998}]{Kubota1998}
{Kubota} A.,  {Tanaka} Y.,  {Makishima} K.,  {Ueda} Y.,  {Dotani} T.,  {Inoue}
  H.,   {Yamaoka} K.,  1998, \mn@doi [\pasj] {10.1093/pasj/50.6.667}, \href
  {https://ui.adsabs.harvard.edu/abs/1998PASJ...50..667K} {50, 667}

\bibitem[\protect\citeauthoryear{{Langmeier}, {Sztajno}, {Truemper}  \&
  {Hasinger}}{{Langmeier} et~al.}{1985}]{Langmeier1985}
{Langmeier} A.,  {Sztajno} M.,  {Truemper} J.,   {Hasinger} G.,  1985, \mn@doi
  [Space Sci. Rev.] {10.1007/BF00179842}, \href
  {https://ui.adsabs.harvard.edu/abs/1985SSRv...40..367L} {40, 367}

\bibitem[\protect\citeauthoryear{{Ludlam} et~al.,}{{Ludlam}
  et~al.}{2022}]{Ludlam2022}
{Ludlam} R.~M.,  et~al., 2022, \mn@doi [\apj] {10.3847/1538-4357/ac5028}, \href
  {https://ui.adsabs.harvard.edu/abs/2022ApJ...927..112L} {927, 112}

\bibitem[\protect\citeauthoryear{{Mainardi} et~al.,}{{Mainardi}
  et~al.}{2010}]{Mainardi2010}
{Mainardi} L.~I.,  et~al., 2010, \mn@doi [A\&A] {10.1051/0004-6361/200912921},
  \href {https://ui.adsabs.harvard.edu/abs/2010A&A...512A..57M} {512, A57}

\bibitem[\protect\citeauthoryear{{Misra} et~al.,}{{Misra}
  et~al.}{2017}]{Misra2017}
{Misra} R.,  et~al., 2017, \mn@doi [\apj] {10.3847/1538-4357/835/2/195}, \href
  {https://ui.adsabs.harvard.edu/abs/2017ApJ...835..195M} {835, 195}

\bibitem[\protect\citeauthoryear{{Mitsuda} et~al.,}{{Mitsuda}
  et~al.}{1984}]{Mitsuda1984}
{Mitsuda} K.,  et~al., 1984, \pasj, \href
  {https://ui.adsabs.harvard.edu/abs/1984PASJ...36..741M} {36, 741}

\bibitem[\protect\citeauthoryear{{Mondal}, {Dewangan}, {Pahari}  \&
  {Raychaudhuri}}{{Mondal} et~al.}{2018}]{Mondal2018}
{Mondal} A.~S.,  {Dewangan} G.~C.,  {Pahari} M.,   {Raychaudhuri} B.,  2018,
  \mn@doi [\mnras] {10.1093/mnras/stx2931}, \href
  {https://ui.adsabs.harvard.edu/abs/2018MNRAS.474.2064M} {474, 2064}

\bibitem[\protect\citeauthoryear{{Mondal}, {Dewangan}  \&
  {Raychaudhuri}}{{Mondal} et~al.}{2019}]{Mondal2019}
{Mondal} A.~S.,  {Dewangan} G.~C.,   {Raychaudhuri} B.,  2019, \mn@doi [\mnras]
  {10.1093/mnras/stz1658}, \href
  {https://ui.adsabs.harvard.edu/abs/2019MNRAS.487.5441M} {487, 5441}

\bibitem[\protect\citeauthoryear{{Nied{\'z}wiecki}, {Szanecki}  \&
  {Zdziarski}}{{Nied{\'z}wiecki} et~al.}{2019}]{Nied2019}
{Nied{\'z}wiecki} A.,  {Szanecki} M.,   {Zdziarski} A.~A.,  2019, \mn@doi
  [\mnras] {10.1093/mnras/stz487}, \href
  {https://ui.adsabs.harvard.edu/abs/2019MNRAS.485.2942N} {485, 2942}

\bibitem[\protect\citeauthoryear{{Nowak}, {Vaughan}, {Wilms}, {Dove}  \&
  {Begelman}}{{Nowak} et~al.}{1999}]{Nowak1999}
{Nowak} M.~A.,  {Vaughan} B.~A.,  {Wilms} J.,  {Dove} J.~B.,   {Begelman}
  M.~C.,  1999, \mn@doi [\apj] {10.1086/306610}, \href
  {https://ui.adsabs.harvard.edu/abs/1999ApJ...510..874N} {510, 874}

\bibitem[\protect\citeauthoryear{{Popham} \& {Sunyaev}}{{Popham} \&
  {Sunyaev}}{2001}]{Popham2001}
{Popham} R.,  {Sunyaev} R.,  2001, \mn@doi [\apj] {10.1086/318336}, \href
  {https://ui.adsabs.harvard.edu/abs/2001ApJ...547..355P} {547, 355}

\bibitem[\protect\citeauthoryear{{Revnivtsev} \& {Gilfanov}}{{Revnivtsev} \&
  {Gilfanov}}{2006}]{Revnivtsev2006}
{Revnivtsev} M.~G.,  {Gilfanov} M.~R.,  2006, \mn@doi [\aap]
  {10.1051/0004-6361:20053964}, \href
  {https://ui.adsabs.harvard.edu/abs/2006A&A...453..253R} {453, 253}

\bibitem[\protect\citeauthoryear{{Saavedra}, {Garc{\'\i}a}, {Fogantini},
  {M{\'e}ndez}, {Combi}, {Luque-Escamilla}  \& {Mart{\'\i}}}{{Saavedra}
  et~al.}{2023}]{Saavedra2023}
{Saavedra} E.~A.,  {Garc{\'\i}a} F.,  {Fogantini} F.~A.,  {M{\'e}ndez} M.,
  {Combi} J.~A.,  {Luque-Escamilla} P.~L.,   {Mart{\'\i}} J.,  2023, \mn@doi
  [\mnras] {10.1093/mnras/stad1157}, \href
  {https://ui.adsabs.harvard.edu/abs/2023MNRAS.522.3367S} {522, 3367}

\bibitem[\protect\citeauthoryear{{Schultheis} et~al.,}{{Schultheis}
  et~al.}{2014}]{Schultheis2014}
{Schultheis} M.,  et~al., 2014, \mn@doi [A\&A] {10.1051/0004-6361/201322788},
  \href {https://ui.adsabs.harvard.edu/abs/2014A&A...566A.120S} {566, A120}

\bibitem[\protect\citeauthoryear{{Schulz}, {Hasinger}  \& {Truemper}}{{Schulz}
  et~al.}{1989}]{Schulz1989}
{Schulz} N.~S.,  {Hasinger} G.,   {Truemper} J.,  1989, \mn@doi [A\&A]
  {10.1007/978-94-009-2273-0_25}, \href
  {https://ui.adsabs.harvard.edu/abs/1989A&A...225...48S} {225, 48}

\bibitem[\protect\citeauthoryear{{Schulz}, {Corrales}  \& {Canizares}}{{Schulz}
  et~al.}{2016}]{Schulz2016}
{Schulz} N.~S.,  {Corrales} L.,   {Canizares} C.~R.,  2016, \mn@doi [\apj]
  {10.3847/0004-637X/827/1/49}, \href
  {https://ui.adsabs.harvard.edu/abs/2016ApJ...827...49S} {827, 49}

\bibitem[\protect\citeauthoryear{{Shimura} \& {Takahara}}{{Shimura} \&
  {Takahara}}{1995}]{Shimura1995}
{Shimura} T.,  {Takahara} F.,  1995, \mn@doi [\apj] {10.1086/175740}, \href
  {https://ui.adsabs.harvard.edu/abs/1995ApJ...445..780S} {445, 780}

\bibitem[\protect\citeauthoryear{{Singh} et~al.,}{{Singh}
  et~al.}{2016}]{Singh2016}
{Singh} K.~P.,  et~al., 2016, in {den Herder} J.-W.~A.,  {Takahashi} T.,
  {Bautz} M.,  eds,  Society of Photo-Optical Instrumentation Engineers (SPIE)
  Conference Series Vol. 9905, Space Telescopes and Instrumentation 2016:
  Ultraviolet to Gamma Ray. p. 99051E, \mn@doi{10.1117/12.2235309}

\bibitem[\protect\citeauthoryear{{Steiner}, {Heinke}, {Bogdanov}, {Li}, {Ho},
  {Bahramian}  \& {Han}}{{Steiner} et~al.}{2018}]{Steiner2018}
{Steiner} A.~W.,  {Heinke} C.~O.,  {Bogdanov} S.,  {Li} C.~K.,  {Ho} W.~C.~G.,
  {Bahramian} A.,   {Han} S.,  2018, \mn@doi [\mnras] {10.1093/mnras/sty215},
  \href {https://ui.adsabs.harvard.edu/abs/2018MNRAS.476..421S} {476, 421}

\bibitem[\protect\citeauthoryear{{Thomas}, {Gudennavar}, {Misra}  \&
  {Bubbly}}{{Thomas} et~al.}{2022}]{Thomas2022}
{Thomas} N.~T.,  {Gudennavar} S.~B.,  {Misra} R.,   {Bubbly} S.~G.,  2022,
  \mn@doi [\apj] {10.3847/1538-4357/ac425e}, \href
  {https://ui.adsabs.harvard.edu/abs/2022ApJ...925..167T} {925, 167}

\bibitem[\protect\citeauthoryear{{Thomas}, {Gudennavar}  \& {Bubbly}}{{Thomas}
  et~al.}{2023}]{Thomas2023}
{Thomas} N.~T.,  {Gudennavar} S.~B.,   {Bubbly} S.~G.,  2023, \mn@doi [\mnras]
  {10.1093/mnras/stad555}, \href
  {https://ui.adsabs.harvard.edu/abs/2023MNRAS.521..433T} {521, 433}

\bibitem[\protect\citeauthoryear{{White}, {Stella}  \& {Parmar}}{{White}
  et~al.}{1988}]{White1988}
{White} N.~E.,  {Stella} L.,   {Parmar} A.~N.,  1988, \mn@doi [\apj]
  {10.1086/165901}, \href
  {https://ui.adsabs.harvard.edu/abs/1988ApJ...324..363W} {324, 363}

\bibitem[\protect\citeauthoryear{{Wijnands}, {van der Klis}  \& {van
  Paradijs}}{{Wijnands} et~al.}{1998}]{Wijnands1998}
{Wijnands} R.,  {van der Klis} M.,   {van Paradijs} J.,  1998, in {Koyama} K.,
  {Kitamoto} S.,   {Itoh} M.,  eds,  International Astronomical Union Series
  Vol. 188, The Hot Universe. p.~370, \mn@doi{10.1007/978-94-011-4970-9_129}

\bibitem[\protect\citeauthoryear{{Wilms}, {Allen}  \& {McCray}}{{Wilms}
  et~al.}{2000}]{Wilms2000}
{Wilms} J.,  {Allen} A.,   {McCray} R.,  2000, \mn@doi [\apj] {10.1086/317016},
  \href {https://ui.adsabs.harvard.edu/abs/2000ApJ...542..914W} {542, 914}

\bibitem[\protect\citeauthoryear{{Yao} et~al.,}{{Yao} et~al.}{2021}]{Yao2021}
{Yao} Y.,  et~al., 2021, \mn@doi [\apj] {10.3847/1538-4357/ac15f8}, \href
  {https://ui.adsabs.harvard.edu/abs/2021ApJ...920..121Y} {920, 121}

\bibitem[\protect\citeauthoryear{{Zdziarski}, {Szanecki}, {Poutanen},
  {Gierli{\'n}ski}  \& {Biernacki}}{{Zdziarski} et~al.}{2020}]{Zdziarski2020}
{Zdziarski} A.~A.,  {Szanecki} M.,  {Poutanen} J.,  {Gierli{\'n}ski} M.,
  {Biernacki} P.,  2020, \mn@doi [\mnras] {10.1093/mnras/staa159}, \href
  {https://ui.adsabs.harvard.edu/abs/2020MNRAS.492.5234Z} {492, 5234}

\bibitem[\protect\citeauthoryear{{{\.Z}ycki}, {Done}  \& {Smith}}{{{\.Z}ycki}
  et~al.}{1999}]{Zycki1999}
{{\.Z}ycki} P.~T.,  {Done} C.,   {Smith} D.~A.,  1999, \mn@doi [\mnras]
  {10.1046/j.1365-8711.1999.02885.x}, \href
  {https://ui.adsabs.harvard.edu/abs/1999MNRAS.309..561Z} {309, 561}

\bibitem[\protect\citeauthoryear{van~den Berg \& Homan}{van~den Berg \&
  Homan}{2017}]{vandenberg2017}
van~den Berg M.,  Homan J.,  2017, \mn@doi [\apj] {10.3847/1538-4357/834/1/71},
  \href {https://ui.adsabs.harvard.edu/abs/2017ApJ...834...71V} {834, 71}

\makeatother
\end{thebibliography}
\bibliographystyle{\mnras}
\label{lastpage}
\end{document}